\documentclass[aps,prc,twocolumn,superscriptaddress,showpacs,nofootinbib]{revtex4-1}
\usepackage{graphicx,amsmath,amssymb,bm,color}

\DeclareGraphicsExtensions{ps,eps,ps.gz,frh.eps,ml.eps}
\usepackage[polutonikogreek,english]{babel}
\usepackage{fmtcount} 
\usepackage{graphicx}
\usepackage{amsmath, amsfonts, amssymb, amsthm, amscd, amsbsy, amstext} 
\usepackage{marvosym} 
\usepackage{yfonts}  
\usepackage{braket}
\usepackage{mathrsfs} 
\usepackage{mathcomp} 
\usepackage{upgreek}
\usepackage{array}

\usepackage{url}
\usepackage{hyperref}%

\DeclareMathOperator{\tr}{\operatorname{tr}}

\newcommand{\breq}{\nonumber \\}

\def\Xint#1{\mathchoice
{\XXint\displaystyle\textstyle{#1}}%
{\XXint\textstyle\scriptstyle{#1}}%
{\XXint\scriptstyle\scriptscriptstyle{#1}}%
{\XXint\scriptscriptstyle\scriptscriptstyle{#1}}%
\!\int}
\def\XXint#1#2#3{{\setbox0=\hbox{$#1{#2#3}{\int}$ }
\vcenter{\hbox{$#2#3$ }}\kern-.6\wd0}}

\def\dashint{\Xint-}

\makeatletter
\newcommand{\rmnum}[1]{\romannumeral #1}
\newcommand{\Rmnum}[1]{\expandafter\@slowromancap\romannumeral #1@}
\makeatother

\makeatletter

\def\env@blscases{
  \let\@ifnextchar\new@ifnextchar
  \left.
  \def\arraystretch{1.2}
  \array{@{}l@{\quad}l@{}}
}
\makeatother

\makeatletter

\def\env@rcases{
  \let\@ifnextchar\new@ifnextchar
  \left.
  \def\arraystretch{1.2}
  \array{@{}l@{\quad}l@{}}
}
\makeatother

\begin {document}
\title{Divergence of the isospin-asymmetry expansion of the nuclear equation of state \\ in many-body perturbation theory}
\author{Corbinian\ Wellenhofer}
\email{corbinian.wellenhofer@tum.de}
\affiliation{Physik Department, Technische Universit\"{a}t M\"{u}nchen, D-85747 Garching, Germany}
\author{Jeremy W.\ Holt}
\email{jwholt.phys@gmail.com}
\affiliation{Department of Physics, University of Washington, Seattle, WA}
\affiliation{Cyclotron Institute and Department of Physics and Astronomy, Texas A\&M University, College Station, TX}
\author{Norbert Kaiser}
\email{n.kaiser@ph.tum.de}
\affiliation{Physik Department, Technische Universit\"{a}t M\"{u}nchen, D-85747 Garching, Germany}

\begin{abstract}
The isospin-asymmetry dependence of the nuclear matter equation of state obtained from microscopic chiral two- and three-body interactions in second-order many-body perturbation theory is examined in detail. The quadratic, quartic and sextic coefficients in the Maclaurin expansion of the free energy per particle of infinite homogeneous nuclear matter with respect to the isospin asymmetry are extracted numerically using finite differences, and the resulting polynomial isospin-asymmetry parametrizations are compared to the full isospin-asymmetry dependence of the free energy. 
It is found that in the low-temperature and high-density regime where the radius of convergence of the expansion is generically zero, the inclusion of higher-order terms beyond the leading quadratic approximation leads overall to a significantly poorer description of the isospin-asymmetry dependence. 
In contrast, at high temperatures and densities well below nuclear saturation density, the interaction contributions to the higher-order coefficients are negligible and the deviations from the quadratic approximation are predominantly from the noninteracting term in the many-body perturbation series.
Furthermore, we extract the leading logarithmic term in the isospin-asymmetry expansion of the equation of state at zero temperature from the analysis of linear combinations of finite differences. It is shown that the logarithmic term leads to a considerably improved description of the isospin-asymmetry dependence at zero temperature.
\end{abstract}

\maketitle

\section{Introduction}\label{sec0}
The isospin-asymmetry dependence of the nuclear thermodynamic equation of state (EoS) is essential for
interpreting data from intermediate-energy heavy-ion collisions and for understanding various nuclear astrophysical 
phenomena (for recent reviews, see Refs.\ \cite{Li:2008gp,Steiner:2004fi}). Assuming charge symmetry 
of the strong interaction \cite{MILLER19901,Epelbaum:1999zn}, previous parametrizations of the
isospin-asymmetry dependence of the free energy per particle $F(T,\rho,\delta)$ of infinite homogeneous nuclear 
matter have employed the Maclaurin expansion
\begin{align} \label{Fexpan1}
F(T,\rho,\delta) \simeq
 \sum_{n=0}^{N} A_{2n}(T,\rho)\, \delta^{2n} =: F_{[2N]}(T,\rho,\delta),
\end{align}
where $T$ is the temperature, $\rho=\rho_\text{n}+\rho_\text{p}$ is the total nucleon density, and
$\delta=(\rho_\text{n}-\rho_\text{p})/ \rho$ is the isospin asymmetry with $\rho_\text{n/p}$ the 
neutron/proton density.
In Eq.\ (\ref{Fexpan1}) the different expansion (Maclaurin) coefficients are given by
\begin{align} \label{Fexpan2}
A_{2n}(T,\rho)=\frac{1}{(2n)!}\frac{\partial^{2n} F(T,\rho,\delta)}{\partial \delta^{2n}}\bigg|_{\delta=0}.
\end{align}
The usefulness of this expansion depends on the accuracy of the approximation polynomials 
$F_{[2N]}(T,\rho,\delta)$ for small values of $N$. The importance of terms beyond the leading quadratic order for various 
properties of neutron stars has been stressed in Refs.\ 
\cite{PhysRevC.89.028801,PhysRevC.74.045808,Cai:2011zn}. 
In the present work we explore in detail the convergence behavior of the isospin-asymmetry expansion of the 
nuclear matter EoS at zero and finite temperature employing second-order many-body perturbation theory \cite{Goldstone:1957zz,RevModPhys.39.771,BDDnuclphys7,Kohn:1960zz,Luttinger:1960ua,brout2,horwitz2,paper1}. 
In particular, for astrophysical simulations of core-collapse supernovae and binary neutron-star mergers we examine the accuracy of the quadratic, quartic, and sextic approximations $F_{[2],[4],[6]}(T,\rho,\delta)$ at high temperatures and densities up to around twice saturation density.
\newline
\indent
From global fits to nuclear binding energies, one expects that at zero temperature and densities 
around nuclear saturation density, already the leading-order quadratic ($N=1$) approximation 
provides a good approximation to the exact isospin-asymmetry dependence of $F(T,\rho,\delta)$.
This has been supported by various many-body calculations with microscopic nuclear forces \cite{PhysRevC.44.1892,Zuo:2002sg,zuo99,Drischler:2013iza,PhysRevC.80.045806,PhysRevC.71.014313}.
The quartic Maclaurin coefficient $A_4$ was found to be of small size ($\lesssim 1\,\text{MeV}$) in self-consistent mean field theory calculations employing phenomenological
\cite{PhysRevC.89.028801,PhysRevC.74.045808,Cai:2011zn} and microscopic 
\cite{PhysRevC.89.028801,PhysRevC.57.3488,Carbone:2013cpa} nuclear forces.
Recent work \cite{PhysRevC.91.065201}, however, has shown that perturbative contributions 
beyond the Hartree-Fock level give rise to a quartic coefficient $A_4$ that is in fact singular at zero temperature; i.e., the second-order contribution to $F(T=0,\rho,\delta)$ is not a smooth function of $\delta$, but of differentiability class $C^3$ only. At finite $T$, however, the same contribution is smooth 
($C^\infty$), but at very low temperatures it cannot be analytic ($C^\omega$) since for $T\rightarrow 0$ its higher-order 
Maclaurin coefficients diverge. At very low temperatures the isospin-asymmetry expansion (of the second-order 
perturbative contribution to the EoS) therefore represents an asymptotic expansion with zero radius of convergence.\footnote{A simple example \cite{book1} of a $C^\infty$ function $f(x)$ whose Maclaurin series has zero radius of convergence is $f(x)=\sum_{n=0}^{\infty}\exp(-n) \cos(n^2 x)$.} The question remains whether the radius of convergence becomes finite for higher temperatures, and more importantly in what region of the parameter space isospin-asymmetry parametrizations $F_{[2N]}(T,\rho,\delta)$ beyond the leading quadratic order are useful. Furthermore, using a simple contact interaction in Ref.\ \cite{PhysRevC.91.065201} a different expansion that includes logarithmic terms $\sim \delta^{2n} \ln|\delta|$ of the zero-temperature EoS was identified; however, the applicability of this expansion in the case of realistic nuclear interactions has not yet been studied.

In the present work we investigate these issues using the modern chiral effective field theory ($\chi$EFT) approach to low-momentum nuclear interactions. In $\chi$EFT, nuclear interactions are organized in a systematic expansion that naturally includes multi-nucleon forces. While the low-energy expansion (power counting) can be adapted for finite-density systems \cite{Kaiser2002255,Epelbaum:2008ga,Holt:2013fwa,Oller:2009zt,Lacour:2009ej}, a complementary approach is to use free-space $\chi$EFT interactions in nuclear many-body calculations with the pertinent low-energy constants fitted to scattering observables and properties of light nuclei. This approach amounts to a prediction of nuclear many-body properties from the underlying effective microscopic theory without further fine-tuning. 
The framework allows to estimate theoretical uncertainties (for recent developments see Refs.\ \cite{Furnstahl:2014xsa,Sammarruca:2014zia,Epelbaum15,Drischler:2015eba,Wesolowski:2015fqa}), although calculations with 
complete uncertainty propagation remain a challenge.

In an EFT the ultraviolet momentum cutoff $\Lambda$ is a variable parameter. Employing interactions with suitably low values of $\Lambda$ enables a perturbative calculation of nuclear thermodynamics. Using low-momentum chiral nuclear interactions in many-body perturbation theory, a realistic thermodynamic EoS of infinite homogeneous nuclear matter has been obtained in Refs.\ \cite{paper1,paper2}. 
More specifically, the results for the EoS of isospin-symmetric nuclear matter and the symmetry energy are consistent with empirical constraints, and the EoS of pure neutron matter at low fugacities reproduces approximately the virial expansion \cite{Horowitz:2005zv}.
This motivates a detailed study of the isospin-asymmetry dependence of the EoS from perturbative calculations with chiral nuclear interactions.

In this paper we examine the isospin-asymmetry expansion of the nuclear EoS obtained from the sets of microscopic chiral low-momentum N3LO (next-to-next-to-next-to-leading leading order) two-body and N2LO (next-to-next-to-leading leading order) three-body interactions n3lo414 ($\Lambda\! =\! 414$\!\! MeV) and n3lo450 ($\Lambda=450$\! MeV) constructed in Refs.\ \cite{Entem:2003ft,Coraggio:2014nvaa,Coraggio13} and used in nuclear many-body calculations and astrophysical studies in Refs.\ \cite{paper1,paper2,Rrapaj:2015zba,Rrapaj:2014yba,Holt13,Holt:2015dfa,Coraggio:2014nvaa,Coraggio13}.
In particular, assuming a quadratic dependence on the isospin asymmetry of the interaction contributions to the free energy per particle, in Ref.\ \cite{paper2} a detailed study of the liquid-gas phase transition of isospin-asymmetric nuclear matter was conducted. 
In the present paper we now analyze in detail the isospin-asymmetry dependence of the interaction contributions, and extract the Maclaurin coefficients $A_{\text{2,4,6}}(T,\rho)$ numerically from higher-order finite-difference approximations.
Using the results for $A_{\text{2,4,6}}(T,\rho)$ we then compare the resulting approximation polynomials to the full isospin-asymmetry dependence of the EoS. 
Since the different many-body contributions behave differently with respect to the isospin-asymmetry expansion, we examine the Maclaurin coefficients for each contribution individually. 

The paper is organized as follows. Sec.\ \ref{sec2} provides details of the calculations and an examination of the Maclaurin expansion of the individual many-body contributions. In addition, in Sec.\ \ref{sec2} we extract the leading logarithmic term in the isospin-asymmetry expansion of the zero-temperature EoS. In Sec.\ \ref{sec1} we discuss the results obtained for the quadratic, quartic and sextic coefficients. 
In Sec.\ \ref{sec3} we then compare the full isospin-asymmetry dependent free energy to various approximations. Finally, Sec.\ \ref{sec4} provides a summary of the paper.

\section{Details of the Calculations}\label{sec2}

In this section we examine the isospin-asymmetry expansion of the
noninteracting contribution\footnote{For the noninteracting contributions see also Ref.\ \cite{paper2}, which contains the
following typos: in the second and third lines of Eq.\ (22) a factor $1/2$ is missing, the expressions given in 
Sec.\ VA for the $\rho\rightarrow 0$ limits should read ``$\bar{F}_\text{rel}\rightarrow - 15 T^2/(8M)$'' and 
``$\bar{E}_\text{rel}\rightarrow 15 T^2/(8M)$'', and the expressions for $L$ and $K_\text{sym}$ given in Sec.\ IV need additional factors $3 \rho_\text{sat}$ and $9 \rho_\text{sat}^2$, respectively.} and the interaction contributions to the nuclear matter EoS.
For the computation of the Maclaurin coefficients corresponding to the interaction contributions, finite-difference approximations have been used. We
discuss the accuracy of the finite-difference method and (at zero temperature) benchmark against the analytic results for the second-order contribution with an $S$-wave contact interaction. The leading logarithmic term in the expansion of the second-order contribution at zero temperature is extracted from the analysis of linear combinations of finite differences.

\subsection{Definitions}\label{sec21}
The symmetry free energy is defined as the difference between the free energy per particle of pure neutron matter ($\delta=1$) and the free energy per particle of (homogeneous) isospin-symmetric nuclear matter ($\delta=0$):
\begin{align} \label{Fexpan3}
F_{\text{sym}}(T,\rho)=F(T,\rho,1)-F(T,\rho,0).
\end{align}
The accuracy
of the leading-order quadratic isospin-asymmetry approximation is inversely related to the magnitude of $F_{\text{sym}}-A_{2}$.
In addition, we consider the function $\xi(T,\rho)$ defined as
\begin{align} \label{Fexpan4}
\xi(T,\rho):=&1-\frac{A_{2}(T,\rho)}{F_{\text{sym}}(T,\rho)},
\end{align}
as well as the functions $\zeta_{2N}(T,\rho)$ defined as
\begin{align} \label{Fexpan4}
\zeta_{2N}(T,\rho):=\frac{\sum_{n=2}^{N}A_{2n}(T,\rho)}{F_{\text{sym}}(T,\rho)}.
\end{align}
If the isospin-asymmetry expansion converges for $\delta \in [-1,1]$ in a given region in the temperature-density plane, then $\zeta_{2N}\xrightarrow{N \rightarrow \infty} \xi$ in that region.

\subsection{Noninteracting and nonrelativistic two-species Fermi gas}\label{sec22}

For a nonrelativistic and noninteracting Fermi gas with two species, $\xi$ and $\zeta_{2N}$ have
parameter-independent limiting values
\begin{align}
\xi |_{\rho \neq 0,T\rightarrow 0}=& 1-10/[9(2^{5/3}-2)]\simeq 0.054, \\
\xi |_{T\rightarrow \infty}=&\xi |_{T\neq 0,\rho\rightarrow 0}= 1-1/\ln(4)\simeq 0.279, \\ \label{zetafree1}
\zeta_{2N}|_{\rho \neq 0,T\rightarrow 0}=& 
\frac{10}{9(2^{5/3}-2)} \sum_{n=2}^{N}\prod_{k=0}^{2n-3}\frac{1+3k}{9+3k},
\\ \label{zetafree2}
\zeta_{2N}|_{T\rightarrow \infty}=&\zeta_{2N}|_{T\neq 0,\rho\rightarrow 0}
=\frac{1}{\ln(2)}\sum_{n=3}^{2N}\frac{(-1)^{n+1}}{n},
\end{align}
where the $T\rightarrow \infty$ and $\rho\rightarrow 0$ limits follow from the asymptotic behavior of polylogarithms.
Furthermore, the ratios of successive Maclaurin coefficients have the limits
\begin{align}
\frac{A_{2n}}{A_{2(n+1)}}\bigg|_{\rho \neq 0,T\rightarrow 0}=&\frac{18n^2+27n+9}{18n^2-21n+5}\xrightarrow{n\rightarrow \infty}1, \\
\frac{A_{2n}}{A_{2(n+1)}}\bigg|_{T\rightarrow \infty}=&
\frac{A_{2n}}{A_{2(n+1)}}\bigg|_{T\neq 0,\rho\rightarrow 0}
\breq 
=&\frac{2n^2+3n+1}{2n^2-n}\xrightarrow{n\rightarrow \infty}1,
\end{align}
i.e., 
the radius of convergence of the Maclaurin expansion of the noninteracting EoS is $R_\delta=1$.

Comparing $\zeta_{2N}/\xi\,|_{\rho \neq 0,T\rightarrow 0}\in$ \{0.646, 0.814, 0.883, 0.918, 0.999\}
and
$\zeta_{2N}/\xi\,|_{T\rightarrow \infty} = \zeta_{2N}/\xi\, |_{T\neq 0,\rho\rightarrow 0}
\in$ \mbox{\{ 0.431, 0.604, 0.696, 0.754, 0.987 \}} for  \mbox{$N \in \{2, 3, 4, 5, 100\}$}, one can deduce that the isospin-asymmetry expansion of the free Fermi gas EoS becomes less accurate (at a fixed order $N$) for increasing values of $T$ and decreasing values of $\rho$. 
In particular, in Ref.\ \cite{paper2} it was shown that the ratios
$A_{2n}/A_{2(n+1)}$ are at finite density monotonic decreasing functions of temperature and at finite temperature monotonic increasing functions of density, and the $T\rightarrow \infty$ limiting values are approached relatively quickly as the temperature is increased. 

\subsection{Finite-difference methods}\label{sec23}

The Maclaurin coefficients can in principle be computed from the explicit expressions obtained for the isospin-asymmetry derivatives (at fixed density and temperature) of the different many-body contributions. 
The length of these expressions however increases  rapidly with the order of the derivative. 
To avoid the numerical evaluation of these lengthy expressions we instead extract the isospin-asymmetry derivatives numerically using finite differences.
The general form of the $\mathscr{N}$-point central finite-difference approximation for $A_{2n}(T,\rho)$ is (using a uniform grid with stepsize $\Delta \delta$ and grid length $N$)
\begin{align} \label{findiff}
A_{2n}(T,\rho)&\simeq A^{N,\Delta \delta}_{2n}(T,\rho)
\breq &=
\frac{1}{(2n)! (\Delta \delta)^{2n}} \!
\sum_{k=0}^{N} (2-\updelta_{k,0})\, \omega_{2n}^{N,k}  F(T,\rho,k \Delta\delta),
\end{align}
where $\mathscr{N}=2N+1\geq 2n+1$.
The finite-difference coefficients $\omega_{2n}^{N,k}$ can be determined by the algorithm given in Ref.\ \cite{coeff}. The formal order of accuracy of the finite-difference approximation $A^{N,\Delta \delta}_{2n}(T,\rho)$ for $A_{2n}(T,\rho)$ is $\Delta \delta^{2N-2n+2}$.
Because $F(T,\rho,\delta)$ can be calculated only to a finite accuracy, $\Delta \delta$ cannot be chosen arbitrarily small without the results being affected by numerical noise.  
Varying $N$ and $\Delta \delta$ provides a means to test the validity of the results for $A_{2n}(T,\rho)$. If the finite-difference approximation is valid, the result should not change under (moderate) variations of $N$ and $\Delta \delta$. 
Since the size of the higher-order derivatives as well as the numerical accuracy varies with the respective many-body contribution as well as the values of the external parameters $T$ and $\rho$, this variation needs to be carried out for every individual contribution and for every single EoS point. 
Carrying out this procedure and systematically increasing the precision of the numerical integration routine in the process, we were able to obtain accurate results for the quadratic, quartic, sextic and to a slightly lesser degree of precision also the octic Maclaurin coefficients. 
In addition, we have extracted the higher-order Maclaurin coefficients also by applying the finite-difference method iteratively, i.e., by  evaluating finite differences of 
\begin{align}
\frac{\partial^n F(T,\rho,\delta)}{\partial \delta^n}\simeq
\frac{1}{(\Delta \delta)^{n}}
\sum_{k=-N}^{N} \omega_{n}^{N,k} \, F(T,\rho,\delta+k \Delta\delta).
\end{align} 
The iterative method involves variable stepsizes and grid lengths at every iteration step and thus behaves differently concerning error systematics; for adequate values of $\Delta \delta$ and $N$ the iterative method gives matching results.
Moreover, as an additional check we have extracted the Maclaurin coefficients for the (total) free energy per particle $F(T,\rho,\delta)$ both by applying finite differences to the data $F(T,\rho,\delta)$ and by summing the Maclaurin coefficients obtained for the individual many-body contributions. 
Finally, we note that in the case of the Hartree-Fock contribution from chiral N2LO three-body  forces at zero temperature, semi-analytical expressions for the Maclaurin coefficients can be derived; the results for the sextic coefficient obtained in this way were found to match the results predicted by the finite-difference method up to four relevant digits.

\subsection{Hartree-Fock level}\label{sec24}

In the following we examine the numerical results for 
$ A_{2,4,6}(T,\rho)$ associated with the first-order contributions from two- 
and three-nucleon interactions. We will find that at the Hartree-Fock level the Maclaurin coefficients are 
hierarchically ordered, $ A_2> A_4> A_6 \hspace{.05in} ( \hspace{.05in} >  A_8)$, and accordingly $\zeta_{4}\simeq \zeta_{6} \simeq \xi$, which indicates that the 
isospin-asymmetry expansion converges at this level. 

\textit{Two-nucleon (NN) interaction.}
The first-order two-body contribution (expanded in partial waves) is given by
\begin{align}
F_{1,\text{NN}}(T,\mu_0^\text{n},\mu_0^\text{p})=&\rho^{-1} \frac{2}{\pi^3}\!  \int \limits_0^{\infty} \! dp \, p^2 
\!\! \int  \limits_0^{\infty}\!\! dK \, K^2  \! \int \limits_{-1}^{1} \! d\cos \theta \nonumber \\
 & \times
\!\sum \limits_{J,\ell,S,t_z} (2J+1) \sum \limits_{\tau_1\leq \tau_2}
\updelta_{t_z,\tau_1+\tau_2}
 \nonumber \\
 & \times
n^{\tau_1}_{|\vec{K}+\vec{p}\,|} n^{\tau_2}_{|\vec{K}-\vec{p}\,|}
\braket{ p | \bar{V}_\text{NN}^{J,\ell,\ell,S,\mathcal{T},t_z}  | p },
\end{align}
where $\theta$ is the angle between $\vec p$ and $\vec K$, and $\bar{V}_{\text{NN}}^{J,\ell_1,\ell_2,S,\mathcal{T},t_z} $ denotes the matrix element of the antisymmetrized NN potential with 
respect to partial-wave states $\Ket{J \ell_i S \mathcal{T} t_z}$.
The contributions from the neutron-neutron (nn), proton-proton (pp) and neutron-proton (np) channels are given by the sum over isospin indices $\tau_1$ and $\tau_2$.
The parameters $\mu_0^\text{n/p}$ (called effective one-body chemical potentials \cite{paper1,Kohn:1960zz,brout2}) are in one-to-one correspondence with the particle densities via
\begin{align} 
\rho_\text{n/p}(T,\mu_0^\text{n/p})=-\upalpha \, T^{3/2} \text{Li}_{3/2}(x_\text{n/p}),
\end{align}
where $x_{\text{n/p}}=-\exp(\mu_0^{\text{n/p}}/T)$, $\upalpha=2^{-1/2}(M/\pi)^{3/2} $ and $\text{Li}_{\nu}(x)$ is the polylogarithm of order $\nu$.

\begin{figure*} 
\centering 
\vspace*{-0.2cm}
\hspace*{-0.0cm}
\includegraphics[width=0.98\textwidth]{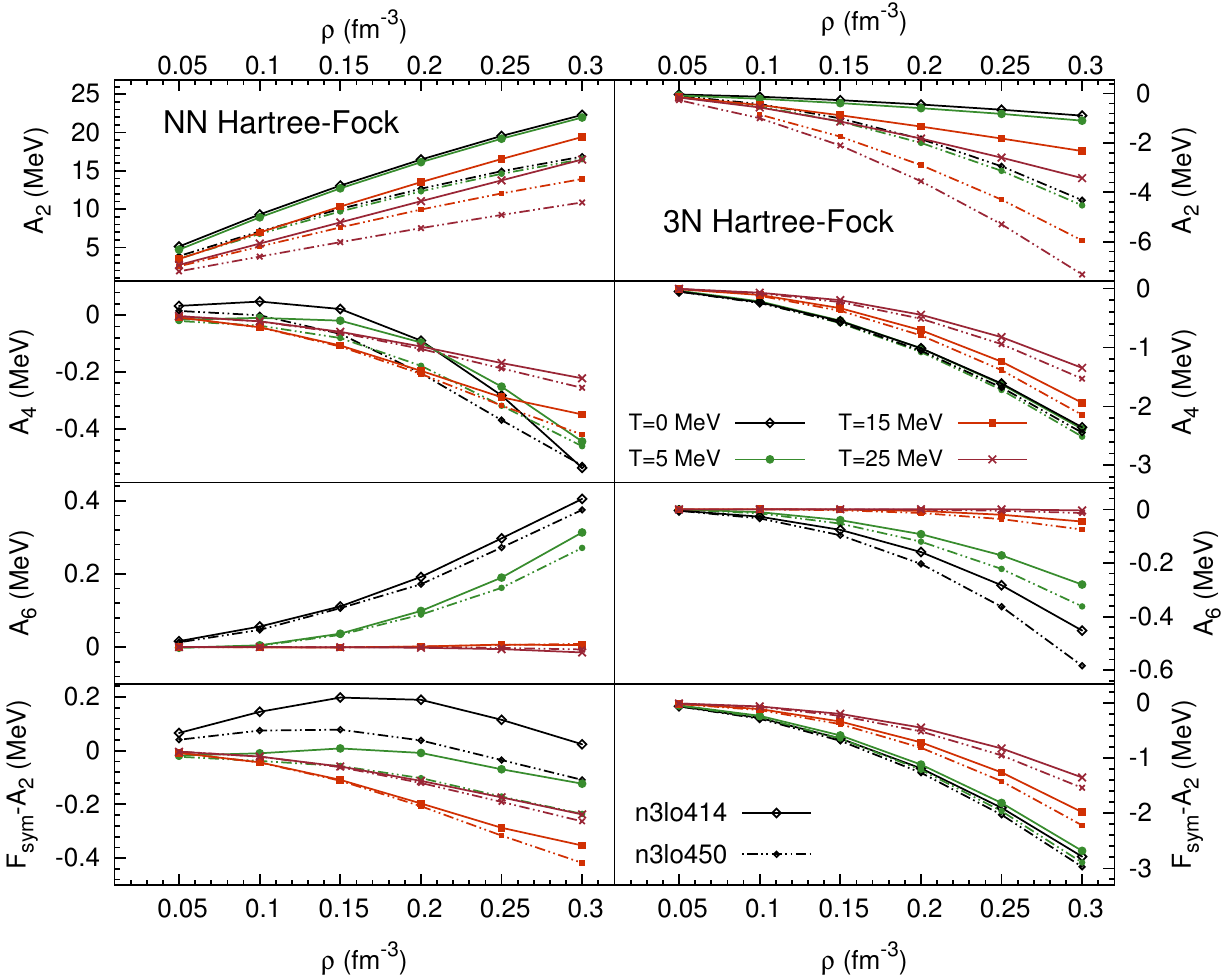} 
\vspace*{-0.0cm}
\caption{(Color online) Density and temperature dependence of the quadratic, quartic and sextic Maclaurin coefficients for the first-order contributions from two- and three-body chiral forces (solid lines for n3lo414, dash-dot lines for n3lo450). Also shown is the difference between the quadratic Maclaurin coefficient and the symmetry free energy.}
\label{plotHFNN1}
\end{figure*}

The results for the quadratic, quartic and sextic Maclaurin coefficients of $F_{1,\text{NN}}(T,\rho,\delta)$ are displayed in the left column of Fig.\ \ref{plotHFNN1}. Also shown are the results for $F_\text{sym}-A_2$.
One sees that the quadratic coefficient $A_2$ greatly outweighs the higher-order coefficients and matches the symmetry free energy $F_\text{sym}$ with high accuracy.
Except for the quartic coefficient $A_4$ the Maclaurin coefficients are monotonic increasing functions of density and decreasing functions of temperature. In the high-temperature regime the Maclaurin coefficients are hierarchically ordered, $A_{2}\gg  A_{4}\gg  A_6$, but this behavior breaks down to some extent at low temperatures where $A_{4}$ and $A_{6}$ are of similar size. Note that the deviations between the n3lo414 and n3lo450 results are significantly reduced in the case of $A_4$ and $A_6$ as compared to $A_2$.

\textit{Three-nucleon (3N) interaction.}
The first-order contribution arising 
from chiral N2LO three-nucleon forces can be written in the compact form \cite{paper1}:
\protect\footnotemark
\begin{align}  \label{3N3bodyInt}
F_{\text{1,3N}}(T,\mu_0^\text{n},\mu_0^\text{p})=&\rho^{-1}
\int \limits_0^{\infty} d k_1 \, \frac{k_1}{2 \pi^2} 
\int \limits_0^{\infty} d k_2 \, \frac{k_2}{2 \pi^2}
\int \limits_0^{\infty} d k_3 \, \frac{k_3}{2 \pi^2} \breq &\times
\mathcal{X}(k_1,k_2,k_3) \;,
\end{align}
where $\mathcal{X}(k_1,k_2,k_3)= \mathcal{X}^{(c_E)} +\mathcal{X}^{(c_D)} +\mathcal{X}^{(\text{Hartree})} +
\mathcal{X}^{(\text{Fock})}$ and the different contributions $\mathcal{X}^{(c_E)}$, $\mathcal{X}^{(c_D)}$, $\mathcal{X}^{(\text{Hartree})}$ and $\mathcal{X}^{(\text{Fock})}$ correspond to the different components of the three-body force (see Fig.\ 2 in Ref.\ \cite{paper1} for the diagrammatic contributions). The explicit expressions for $\mathcal{X}^{(c_E)}$, $\mathcal{X}^{(c_D)}$, $\mathcal{X}^{(\text{Hartree})}$ and $\mathcal{X}^{(\text{Fock})}$ are given by
\begin{align}
\mathcal{X}^{(c_E)}=&\frac{1}{2} \mathcal{K}^{(c_E)} 
n^{\mathrm{p}}_{k_1} n^{\mathrm{n}}_{k_2} \left( n^{\mathrm{p}}_{k_3}+n^{\mathrm{n}}_{k_3} \right),
\nonumber \\
\mathcal{X}^{(c_D)}=&\frac{1}{6} \mathcal{K}^{(c_E)} 
\Big[ \left( n^{\mathrm{p}}_{k_1}+ 2 n^{\mathrm{n}}_{k_1} \right) n^{\mathrm{p}}_{k_2} n^{\mathrm{n}}_{k_3}
+ \left( n^{\mathrm{n}}_{k_1}+ 2 n^{\mathrm{p}}_{k_1} \right)
\breq &\times n^{\mathrm{n}}_{k_2} n^{\mathrm{p}}_{k_3} \Big],
\nonumber \\
\mathcal{X}^{(\text{Hartree})}=&\frac{1}{12} \mathcal{K}^{(\text{Hartree})}
\left(  n^{\mathrm{p}}_{k_1} n^{\mathrm{p}}_{k_2}+ 4 n^{\mathrm{p}}_{k_1} n^{\mathrm{n}}_{k_2} + n^{\mathrm{n}}_{k_1} n^{\mathrm{n}}_{p_2} \right) 
\breq &\times \left( n^{\mathrm{p}}_{k_3}+ n^{\mathrm{n}}_{k_3} \right),
 \nonumber \\
\mathcal{X}^{(\text{Fock})}=&
\frac{1}{6} \left( \mathcal{K}^{(\text{Fock},\,c_{1})} +\mathcal{K}^{(\text{Fock},\,c_{3})} \right)
\Big[ \left(  n^{\mathrm{p}}_{k_1} n^{\mathrm{p}}_{k_2} \right.
\breq & \left.+ 2 n^{\mathrm{n}}_{p_1} n^{\mathrm{n}}_{k_2} \right) n^{\mathrm{p}}_{p_3}+
\left( 2 n^{\mathrm{p}}_{k_1} n^{\mathrm{p}}_{k_2} +  n^{\mathrm{n}}_{k_1} n^{\mathrm{n}}_{k_2} \right) n^{\mathrm{n}}_{k_3} \Big] \nonumber \\
&+\frac{1}{6} \mathcal{K}^{(\text{Fock},\,c_{4})}  
\Big[ \left(  n^{\mathrm{p}}_{k_1}  + 2 n^{\mathrm{n}}_{k_1} \right) n^{\mathrm{p}}_{k_2} n^{\mathrm{n}}_{k_3}
\breq &+
\left( 2 n^{\mathrm{p}}_{k_1}  +  n^{\mathrm{n}}_{k_1} \right) n^{\mathrm{n}}_{k_2} n^{\mathrm{p}}_{k_3}\Big],
\end{align}
where the kernels $\mathcal{K}^{(c_E)}$, $ \mathcal{K}^{(c_D)}$ and  $\mathcal{K}^{(\text{Hartree})}$ are given by Eqs.\ $(18)-(20)$ in Ref.\ \cite{paper1}, and $\mathcal{K}^{(\text{Fock},\,c_{i})}$ is the part of the kernel  $\mathcal{K}^{(\text{Fock})}$ given by Eq.\ (21) in Ref.\ \cite{paper1} proportional to the low-energy constant $c_i$.

In the right column of Fig.\ \ref{plotHFNN1} the Maclaurin coefficients for $F_{1,\text{3N}}(T,\rho,\delta)$ are shown, as well as the corresponding results for $F_\text{sym}-A_2$. 
Overall, the temperature and density dependence of the first-order three-body Maclaurin coefficients is similar to the behavior of the first-order two-body results (the temperature and density dependence of the convergence rate is opposite to the behavior
of the noninteracting contribution).
Similar to the two-body results, the n3lo414 and n3lo450 results are quite similar for $A_{4}$ and $A_{6}$ while the $A_{2}$ values differ more significantly. However, $A_{2}$ is relatively small and
$F_\text{sym}-A_2$ is significantly larger for the three-body contribution, in particular for n3lo414 where $A_4>A_2$ at low values of $T$. 
The relatively small size of $A_2$ results from the cancellation of contributions proportional to different low-energy constants (LECs). For different choices of the LECs (for example, Nijmegen LECs, cf.\ Ref.\ \cite{paper1}) the overall 
size of the first-order three-body contribution to $A_2$ (and also to $A_0$) can be increased.

\begin{widetext}
\subsection{Second-order contribution}\label{sec26}

\begin{figure*}
\centering
\vspace*{-0.2cm}
\hspace*{-0.0cm}
\includegraphics[width=0.98\textwidth]{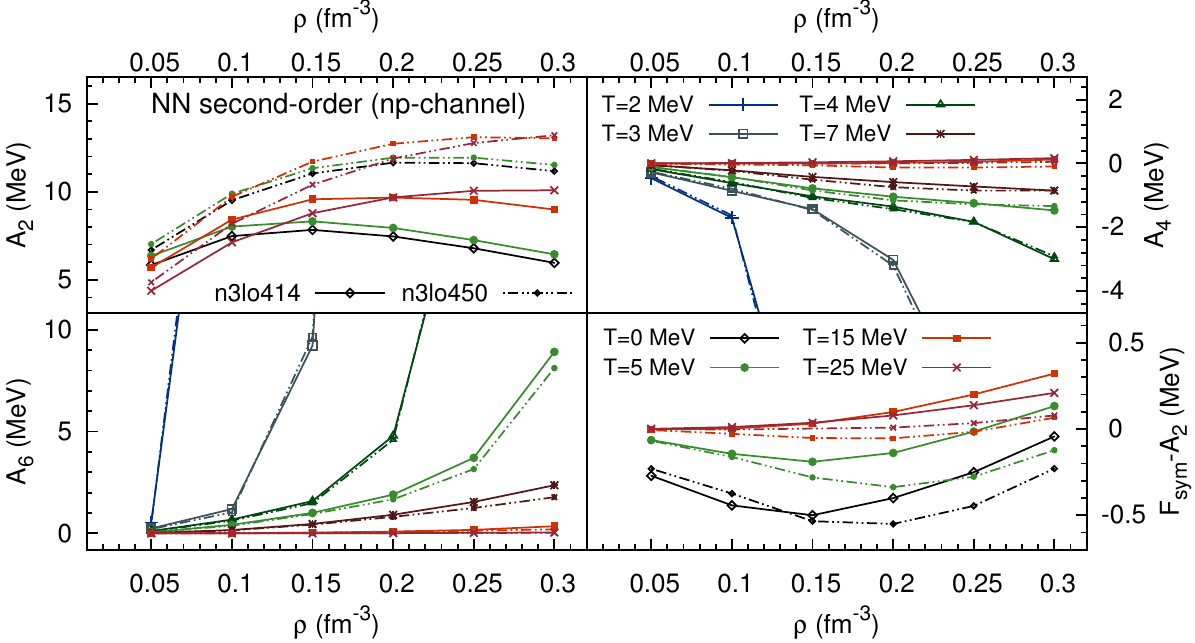} 
\vspace*{-0.0cm}
\caption{(Color online) Quadratic, quartic and sextic Maclaurin coefficients for the np-channel second-order normal contribution (with NN forces only). Also shown are the results for $F_\text{sym}(T,\rho) - A_2(T,\rho)$.}
\label{plot2NN1}
\vspace*{-0.2cm}
\end{figure*}

The partial-wave representation of the second-order (normal) NN contribution to the free energy per particle is given by
\begin{align}  \label{pw2normalT}
F_{2,\text{NN}}(T,\mu_0^\text{n},\mu_0^\text{p})=&-\rho^{-1} \frac{4}{\pi^2}   
\int  \limits_0^{\infty}\!\! d K \, K^2
\int \limits_0^{\infty} \!\! d p_1 \, p_1^2  
\int \limits_0^{\infty} \!\! d p_2 \, p_2^2  
\int \limits_{-1}^{1}\!\! d \cos \theta_1 
\int \limits_{-1}^{1}\!\! d \cos \theta_2   
\sum \limits_{J,\ell_1,\ell_2} \sum \limits_{J',\ell_1',\ell_2'} \sum \limits_{S, t_z} 
\sum \limits_{\tau_1\leq \tau_2} \!\!
\updelta_{t_z,\tau_1+\tau_2} \!\!
\sum \limits_{\tau_3 \leq \tau_4} \!\!
\updelta_{t_z,\tau_3+\tau_4}
\nonumber \\  & \times  
i^{\ell_2-\ell_1}
i^{\ell_1'-\ell_2'}
\sum \limits_{M,m,m^\prime}  \,  {\cal C}(\theta_1,\theta_2)
\;\;\mathcal{G}(p_1,p_2,K,\theta_1,\theta_2) \,
\braket{ p_1 |
\bar{V}_\text{NN}^{J,\ell_1,\ell_2,S,\mathcal{T},t_z}| p_2 }  
\braket{ p_2 |
\bar{V}_\text{NN}^{J'\!\!,\ell_1',\ell_2',S,\mathcal{T},t_z}| p_1 }  .
\end{align}
The function ${\cal C}(\theta_1,\theta_2)$ collects Clebsch-Gordan coefficients and spherical harmonics (see Eq.\ (29) in Ref.\ \cite{paper1}), and the function $\mathcal{G}(p_1,p_2,K,\theta_1,\theta_2)$ is given by

\begin{align} \label{edenominator}
\mathcal{G} &=\frac{
n^{\tau_1}_{|\vec{K}+\vec{p}_1|} n^{\tau_2}_{|\vec{K}-\vec{p}_1|} 
\big(1-{n}^{\tau_3}_{|\vec{K}+\vec{p}_2|}\big)
\big(1-{n}^{\tau_4}_{|\vec{K}-\vec{p}_2|}\big) - 
\big(1-{n}^{\tau_1}_{|\vec{K}+\vec{p}_1|}\big)
\big(1-\bar{n}^{\tau_2}_{|\vec{K}-\vec{p}_1|}\big) 
n^{\tau_3}_{|\vec{K}+\vec{p}_2|} n^{\tau_4}_{|\vec{K}-\vec{p}_2|}}
{\varepsilon(|\vec{K}+\vec{p}_2|,\tau_3)+
\varepsilon(|\vec{K}-\vec{p}_2|,\tau_4)-
\varepsilon(|\vec{K}+\vec{p}_1|,\tau_1)-
\varepsilon(|\vec{K}-\vec{p}_1|,\tau_2)},
\end{align}
\end{widetext}

\noindent
where the single-particle energies are given by $\varepsilon(k,\tau)=k^2/(2M)$.
At finite temperature the integrand in Eq.\ (\ref{pw2normalT}) is smooth, but (for a gapless spectrum) at zero temperature it diverges at the boundary of the integration region.
This is the origin of the singularity of $A_{2n \geq 4}$ at zero temperature.

\footnotetext[3]{Note that the expression for $F_\text{1,3N}$ does not involve a regulator. The low-energy constants $c_{1,3,4}$, which give the dominant contribution to $F_\text{1,3N}$, have been fitted to NN scattering data using nonlocal regulators \cite{Coraggio13} (which can be omitted for $F_\text{1,3N}$, cf.\ Refs.\ \cite{paper1,Dyhdalo:2016ygz}), but the fitting of $c_E$ and $c_D$ was carried out with local regulators \cite{Coraggio:2014nvaa}. The refitting of $c_E$ and $c_D$ employing nonlocal regulators may be warranted \cite{Dyhdalo:2016ygz}; however, the qualitative results for the Maclaurin coefficients for $F_\text{1,3N}$ are not affected by this issue because the $c_E$ and $c_D$ terms give only small contributions to $F_\text{1,3N}$.} 

At zero temperature the Fermi-Dirac distribution functions in Eq.\ (\ref{edenominator}) become step functions, which can be absorbed into the boundaries of the integrals. The first part of the numerator of $\mathcal{G}$ is associated with the following conditions on the angular integrals:
\begin{align} 
&(1.\rmnum{1}) \;\;\;
-\text{min}(\alpha_1^{\tau_2}\,,\,1) \leq \cos \theta_1  \leq \text{min}(\alpha_1^{\tau_1}\,,\,1),\nonumber \\
&(2.\rmnum{1}) \;\;\;
-\text{min}(-\alpha_2^{\tau_3}\,,\,1) \leq \cos \theta_2  \leq \text{min}(-\alpha_2^{\tau_4}\,,\,1),\nonumber
\end{align}
where $\alpha_i^{\tau}=[(k_{\!F}^{\tau})^2-K^2-p_i^2]/(2 K p_i)$, with $k_{\!F}^{\text{n/p}}$ the neutron/proton Fermi momentum.
The integration region for $\theta_1$ vanishes unless the following two conditions are satisfied:
\begin{align} 
&(1.\rmnum{2}) \;\;\; -\alpha_1^{\tau_2}\leq \alpha_1^{\tau_1}
\; \Leftrightarrow  \;
K^2+p_1^2\leq [(k_{\!F}^{\tau_1})^2+(k_{\!F}^{\tau_2})^2]/2 \nonumber \\
&(1.\rmnum{3}) \;\;\; \text{min}(\alpha_1^{\tau_1},\alpha_1^{\tau_2})\geq -1 
\; \Leftrightarrow  \;  \text{min}(k_{\!F}^{\tau_1},k_{\!F}^{\tau_2}) \geq |K-p_1|.\nonumber
\end{align}

\begin{table*}
\begin{center}
\begin{minipage}{14cm}
\setlength{\extrarowheight}{1.5pt}
\hspace*{-13mm}
\begin{tabular}{ ccc c c c c}
\hline  \hline
&\;\;\;\; \;\;\;
$A_{0}\,(\text{MeV})$ &\;\;\;\; \;\;\;$A_{2}\,(\text{MeV})$ &\;\;\;\;\;\;\;$A_{4}\,(\text{MeV})$ &\;\;\;\;\;\;\;
$A_{6}\,(\text{MeV})$&\;\;\;\;\;\;\;
$A_{8}\,(\text{MeV})$&\;\;\;\;\;\;\;$F_{\text{sym}}\,(\text{MeV})$
\\ \hline 
NN (np-channel) &\;\;\;\; \;\;\;
-8.13  &\;\;\;\;\;\;\; 
8.32  &\;\;\;\;\;\;\; 
-0.78  &\;\;\;\;  \;\;\;
 1.01   &\;\;\;\;  \;\;\;
 -0.9   &\;\;\;\;  \;\;\;
8.13 \\ 
NN (total) &\;\;\;\; \;\;\;
-9.60 &\;\;\;\;\;\;\; 
8.64  &\;\;\;\;\;\;\; 
-0.80 &\;\;\;\;  \;\;\;
1.02  &\;\;\;\; \;\;\; 
 -0.9   &\;\;\;\;  \;\;\;
8.36 \\ 
NN+DDNN (total) &\;\;\;\; \;\;\;
-10.80 &\;\;\;\;\;\;\; 
10.48  &\;\;\;\;\;\;\; 
-1.24 &\;\;\;\;  \;\;\;
0.82  &\;\;\;\; \;\;\; 
 -0.8   &\;\;\;\;  \;\;\;
9.63 \\ 
 \hline \hline 
\end{tabular}
\end{minipage}
\begin{minipage}{17.8cm}
\caption
{Different second-order (normal) contributions to the Maclaurin coefficients and the symmetry free energy at $T=5\,\text{MeV}$ and $\rho=0.15\,\text{fm}^{-3}$ (results for n3lo414), see text for details.}
\label{table1}
\end{minipage}
\end{center}
\vspace*{-0.4cm}
\end{table*}

\noindent
Similarly, the integration region for $\theta_2$ vanishes unless the following conditions are satisfied:
\begin{align}
&(2.\rmnum{2}) \;\;\; \alpha_2^{\tau_3}\leq -\alpha_2^{\tau_4}
\;\Leftrightarrow  \
K^2+p_2^2\geq [(k_{\!F}^{\tau_3})^2+(k_{\!F}^{\tau_4})^2]/2,
\nonumber \\
&(2.\rmnum{3}) \;\;\;\text{max}(\alpha_2^{\tau_3},\alpha_2^{\tau_4})\leq 1 
\; \Leftrightarrow  \; \text{max}(k_{\!F}^{\tau_3},k_{\!F}^{\tau_4}) \leq K+p_2.\nonumber
\end{align}
In addition, the following condition arises from the requirement that the intersection of the first two Fermi spheres is nonvanishing:
\begin{align}
&(1.\rmnum{4}) \;\;\; \{K,p_1\}\leq (k_{\!F}^{\tau_1}+ k_{\!F}^{\tau_2})/2. \nonumber
\end{align}
Fixing the order of the momentum integrals as in Eq. (\ref{pw2normalT}) the conditions $1.\rmnum{2}$,$\rmnum{3}$,$\rmnum{4}$ and $2.\rmnum{2}$,$\rmnum{3}$ become
\begin{align}
&(1.\rmnum{4}) \;\;\; 0\leq K \leq (k_{\!F}^{\tau_1}+ k_{\!F}^{\tau_2})/2, 
\nonumber \\
&(1.\rmnum{2},\!\rmnum{3}) \;\;\;  \text{max}[0, K-\text{min}(k_{\!F}^{\tau_1},k_{\!F}^{\tau_2})]
\leq p_1  
\nonumber \\ & \hspace*{3.165cm}
\leq
\text{min}[K+\text{min}(k_{\!F}^{\tau_1},k_{\!F}^{\tau_2}) , \kappa(\tau_1,\tau_2)],
 \nonumber \\
&(2.\rmnum{2},\!\rmnum{3}) \;\;\;  \text{max}[0,\kappa(\tau_3,\tau_4),\text{max}(k_{\!F}^{\tau_3},k_{\!F}^{\tau_4})-K]\leq p_2 < \infty, \nonumber
\end{align}
where $\kappa(\tau_1,\tau_2)=\kappa(\tau_3,\tau_4)=\sqrt{[(k_{\!F}^{\tau_1})^2+(k_{\!F}^{\tau_2})^2]/2-K^2}$. The second part of the numerator of $\mathcal{G}$ leads to the same condition on the integral boundaries, but with $(p_1,\theta_1)$ and $(p_2,\theta_2)$ interchanged. The energy denominator is antisymmetric under $(p_1,\theta_1)\leftrightarrow(p_2,\theta_2)$, therefore both 
parts of $\mathcal{G}$ yield identical contributions.

At second order, the expressions for the contributions involving 3N interactions become more involved. A subset of the diagrams can be simplified by approximating the genuine 3N interactions by a temperature, density, and isospin-asymmetry dependent effective two-body (DDNN) potential $\bar{V}_{\text{DDNN}}(T,\mu_0^\text{n},\mu_0^\text{p})$, cf.\ Refs.\ \cite{Sammarruca:2015tfa,paper1}, which is constructed by closing one nucleon line and integrating over the occupied single-nucleon states (see also Refs.\ \cite{Bogner05,Holt09,Holt:2009ty,PhysRevC.82.014314,Carbone:2014mja,Drischler:2015eba} for further details). In this approximation the total second-order (normal) contribution is given by substituting in Eq.\ (\ref{pw2normalT}) for $\bar{V}_{\text{NN}}$ the quantity $\bar{V}_{\text{total}}(T,\mu_0^\text{n},\mu_0^\text{p})=\bar{V}_{\text{NN}}+\bar{V}_{\text{DDNN}}(T,\mu_0^\text{n},\mu_0^\text{p})$. 

The np-channel results (using the NN potential only) for $A_{2,4,6}(T,\rho)$ and $F_\text{sym}(T,\rho)-A_2(T,\rho)$ are displayed in Fig.\ \ref{plot2NN1}.
Similar to the results for the first-order two-body contribution, $F_\text{sym}-A_2$ is small.
Again the differences between the n3lo414 and n3lo450 results are significantly decreased for $A_{4}$ and $A_{6}$ as compared to $A_2$.
In the high-temperature and low-density region it is $A_2 \gg A_4> A_6 \hspace{.05in} ( \hspace{.05in} > A_8)$, which indicates that the second-order contribution is an analytic function of the isospin asymmetry when the temperature (density) is sufficiently large (small).
At high density and low temperature, however, this behavior breaks down,
and terms beyond $A_2$ diverge with alternating sign in the zero-temperature limit. This observation will be addressed in detail in the following subsection.

Up to now we have focused attention on the np-channel (normal) contributions (from NN forces only) at second order in many-body perturbation theory. In fact, this term 
gives the dominant contribution to the isospin-asymmetry dependence. In Table \ref{table1} we compare the results for different second-order (normal) contributions to $A_{2n}(T,\rho)$, i.e., the np-channel NN contribution, the total NN contribution, and the total contribution with the combined NN and DDNN potential. One sees that for the higher-order Maclaurin coefficients the difference between the total NN and the np-channel NN contribution almost vanishes, which indicates that the nn- and pp-channels are regular also for realistic nuclear interactions.\footnote{Although at very low temperatures we were not able to obtain accurate results for the nn- and pp-channel ${A}_{4,6,8}$ (owing to the structure of ${\cal G}$), we have observed that the finite-difference results are small and do not show an approximately logarithmic stepsize dependence at zero temperature.} 
The deviations between the NN and the NN+DDNN results are more sizable, but the Maclaurin coefficients are still of similar order of magnitude. Because the numerical evaluation of the second-order contribution becomes more involved with the DDNN potential included, in the present paper we have restricted the detailed examination of the isospin-asymmetry expansion at second-order perturbation theory to the (np-channel) NN contribution.

The calculations of Refs.\ \cite{paper1,paper2} include (first-order) self-energy corrections to the single-particle energies appearing in the expression for the second-order (normal) diagram, i.e.,
\begin{align} \label{selfenergyPW}
\varepsilon(k_i,\tau_i; T,\mu_0^\text{n},\mu_0^\text{p}) =&\frac{k_i^2}{2M}+ \tr_{\sigma_j,\tau_j}  \!\! \int  \! \! \frac{\mathrm{d}^3k_j }{(2\pi)^3}\, \,
 n(k_j,\tau_j)  \, \,
\breq & \times  
\braket{  \boldsymbol i \boldsymbol j | \bar{V}_{\text{NN}}+\frac{1}{2}\bar{V}_{\text{DDNN}}|\boldsymbol i \boldsymbol j}.
\end{align}
This is usually approximated as 
\begin{align}		\label{selfenergyPW2}
\varepsilon(k,\tau; T,\mu_0^\text{n},\mu_0^\text{p})\simeq 
\frac{k^2}{2M^{*}(\tau;T,\mu_0^\text{n},\mu_0^\text{p})}+U_0(\tau;T,\mu_0^\text{n},\mu_0^\text{p}).
\end{align}
where $M^{*}$ is called the effective mass. At the Hartree-Fock level, the effective mass is a decreasing function of density and ranges from $0.6-0.7 M$ in isospin-symmetric nuclear matter at saturation density. Higher-order perturbative corrections increase the effective mass close to the bare mass \cite{bertsch68,zuo99,holt11,PhysRevC.82.014314} in the vicinity of the Fermi surface. In the present study, we have not included self-energy insertions into the second-order diagrams, since the global effective mass plus energy shift approximation in Eq.\ (\ref{selfenergyPW2}) does not provide sufficient accuracy when implemented in the extraction of the higher-order Maclaurin coefficients. A more precise treatment including the exact momentum dependence may be warranted.

\subsection{Extraction of leading logarithmic term at zero temperature}\label{sec25}
\textit{Second-order contribution with contact interaction}. To better understand the divergent behavior in the higher-order Maclaurin coefficients seen in the previous section 
for the second-order (normal) diagram, we consider now the simpler case of an $S$-wave contact interaction
$V_{\text{contact}}=\pi M^{-1} (a_s+3 a_t+(a_t-a_s)\vec \sigma_1 \cdot \vec \sigma_2)$. 
For the $S$-wave contact interaction the (dimensionally regularized \cite{Kaiser:2015vpa}) second-order (normal) contribution can be written as
\newpage
\begin{widetext}
\vspace*{-5mm}
\begin{align}
F_{\text{2,contact}}(T,\mu_0^\text{n},\mu_0^\text{p})
=&\rho^{-1}
\frac{1}{2 \pi^4 M} \Big(
a_s^2 \, \Gamma^{\text{nn}}(T,\mu_0^\text{n})
+a_s^2\,\Gamma^{\text{pp}}(T,\mu_0^\text{p}) 
+(3 a_t^2+a_s^2)\,\Gamma^{\text{np}}(T,\mu_0^\text{n},\mu_0^\text{p})
\Big),
\end{align}
where the functions $\Gamma^{\text{nn}/\text{pp}/\text{np}}$ are defined as
\begin{align} \label{Swaveintegral}
\Gamma^{\text{nn}/\text{pp}/\text{np}}=&
  \int_0^{\infty} \!\!\!\! d k_3  
\int_{-1}^{1} \!\!\!\! d y 
\int_0^{\infty} \!\!\!\! d q  \;
\dashint_0^{\infty} \!\!\!\! d k_1 \;
\dashint_{-1}^{1} \!\!\!\! d x \;
\frac{k_3^2 k_1 ^2 q}{k_3 y-k_1 x} \times
\begin{cases}
& \!\!\!\! 2 n_{k_1}^{\text{n}} n_{k_2}^{\text{n}} n_{k_3}^{\text{n}}\\
& \!\!\!\! 2 n_{k_1}^{\text{p}} n_{k_2}^{\text{p}} n_{k_3}^{\text{p}}\\
& \!\!\!\! n_{k_1}^{\text{n}} n_{k_2}^{\text{p}} n_{k_3}^{\text{n}}
+n_{k_1}^{\text{p}} n_{k_2}^{\text{n}} n_{k_3}^{\text{p}}
\end{cases},
\end{align}
with $k_2=(k_3^2-2 k_3 q y + q^2)^{1/2}$. The dashed integral denotes the principal value.
At zero temperature the integrals in Eq.\ (\ref{Swaveintegral}) can be resolved in closed form \cite{PhysRevC.91.065201}, leading to
\begin{align} 
\Gamma^{\text{nn}}(T=0,\rho,\delta)=&
\frac{4 k_{\!F}^7}{105} \big(11 - 2 \ln(2) \big) 
(1 +\delta)^{7/3},   \\
\Gamma^{\text{pp}}(T=0,\rho,\delta)=&
\frac{4 k_{\!F}^7}{105} \big(11 - 2 \ln(2) \big) 
(1 -\delta)^{7/3}, \\ \label{Gammanp}
\Gamma^{\text{np}}(T=0,\rho,\delta)=&\frac{k_{\!F}^7}{420} \sum_\pm
(1 \pm \delta)^{7/3} 
\bigg[
-\frac{8 (1 \mp \delta)^{5/3}}{(1 \pm \delta)^{5/3}} 
+\frac{66 (1 \mp \delta)}{(1 \pm \delta)} 
+\frac{30 (1 \mp \delta)^{1/3}}{(1 \pm \delta)^{1/3}} 
+\bigg(\!\!
-\frac{35 (1 \mp \delta)^{4/3}}{(1 \pm \delta)^{4/3}} 
 \nonumber \\ 
&
+\frac{42 (1 \mp \delta)^{2/3}}{(1 \pm \delta)^{2/3}} 
-15
\bigg)  \ln| K_1|
+\frac{8 (1 \mp \delta)^{7/3}}{(1 \pm \delta)^{7/3}}  \ln|K_2^\mp|
\bigg],
\end{align} 
where $k_{\!F}=(3\pi^2\rho/2)^{1/3}$ is the nucleon Fermi momentum.
The np-channel contribution involves terms proportional to $\ln|K_1|$ and $\ln|K_2^\mp|$, with
$K_1=[(1+\delta)^{1/3}+(1-\delta)^{1/3}]/[(1+\delta)^{1/3}-(1-\delta)^{1/3}]$ and 
$K_2^\mp=(1\mp\delta)^{2/3}/[(1+\delta)^{2/3}-(1-\delta)^{2/3}]$. Both $K_1$ and $K_2^\mp$ exhibit a Laurent series with principal part $\sim\delta^{-1}$. Expanding the logarithms $\ln|K_1|$ and $\ln|K_2^\mp|$ around the principal part and the remaining isospin-asymmetry dependent terms in the expression for the np-channel second-order contribution around $\delta=0$ one obtains a series of the form 
\begin{align} \label{T0expansion}
F(T=0,\rho,\delta)=
A_{0}(T=0,\rho)+ A_{2}(T=0,\rho)\, \delta^{2}+
\sum_{n=2}^\infty A_{2n,\text{reg}}(\rho)\, \delta^{2n}+
\sum_{n=2}^\infty A_{2n,\text{log}}(\rho) \, \delta^{2n} \ln{|\delta|}.
\end{align}
From Eq.\ (\ref{T0expansion}) it follows for the $\delta\rightarrow 0$ limit of the isospin-asymmetry derivatives of $F(T,\rho,\delta)$ at zero temperature:
\begin{align}\label{T0expansion2}
\frac{1}{(2n)!}\frac{\partial^{2n} F(T,\rho,\delta)}{\partial \delta^{2n}}\bigg|_{n\geq2,T=0,\delta \rightarrow 0}
=&
A_{2n,\text{reg}}+A_{2n,\text{log}}\sum_{k=1}^{2n}\frac{1}{k}
+A_{2n,\text{log}}\ln|\delta| -
\sum_{k=2}^{n-1} \frac{(2k)!(2n-2k-1)!}{(2n)!} \frac{ A_{2k,\text{log}} }{\delta^{2(n-k)}}
\bigg|_{\delta \rightarrow 0}
\nonumber \\ 
=&
-\infty \times \text{sign}(A_{4,\text{log}}),
\end{align}
\end{widetext}
i.e., the degree of divergence increases with $n$, and all higher-order derivatives diverge with equal sign. This behavior is impossible for the $T\rightarrow 0$ limit of the higher-order Maclaurin coefficients, i.e., the $T\rightarrow 0$ and the $\delta\rightarrow 0$ limits of the isospin-asymmetry derivatives $\partial^{2(2n+1)} F_2/\partial \delta^{2(2n+1)}$ cannot commute for $n\geq 1$:
\begin{align}
\frac{\partial^{2(2n+1)} F_{\text{2}}}{\partial \delta^{2(2n+1)}}\bigg|_{n\geq1,T=0,\delta\rightarrow 0}
\neq
\frac{\partial^{2(2n+1)} F_{\text{2}}}{\partial \delta^{2(2n+1)}}\bigg|_{n\geq1,\delta=0,T\rightarrow 0}.
\end{align}
This is explained as follows. As can be inferred from Eq.\ (\ref{T0expansion2}), at zero temperature the higher-order isospin-asymmetry derivatives $\partial^{2n} F_2/\partial \delta^{2n}$, for $n\geq 2$, all have positive (isospin-asymmetry) slope and negative (isospin-asymmetry) curvature for $\delta=0+\epsilon$. 
This behavior is impossible at finite temperature where $F_2\in C^\infty$, and hence $\partial^{2n+1}F_2/\partial \delta^{2n+1}=0$ at $\delta=0$ (by charge symmetry). If $\partial^{4}F_2/\partial \delta^{4}$ has positive slope for $T\neq 0$ and $\delta\rightarrow 0$ then it must be convex at $\delta=0$, thus
$\partial^{6} F_2/\partial \delta^{6}$ can only diverge with positive sign and curvature for $\delta=0$ and $T\rightarrow 0$, etc. 

\begin{table*}
\vspace*{-0.45cm}
\begin{center}
\begin{minipage}{14cm}
\setlength{\extrarowheight}{1.5pt}
\hspace*{-18mm}
\begin{tabular}{ ccc c c c c}
\hline  \hline
$\rho/{\text{fm}^{-3}}$&\;\;\;\;
$0.05$ &\;\;\;\;
$0.10$ &\;\;\;\;
$0.15$&\;\;\;\;
$0.20$ &\;\;\;\;
$0.25$&\;\;\;\;
$0.30$
\\ \hline 
$A_{2}(T=0,\rho)/\text{MeV}$ &\;\;\;\;
16.48 (16.03) &\;\;\;\;
24.41 (24.92) &\;\;\;\;
32.03 (31.41) &\;\;\;\;
36.94 (36.08) &\;\;\;\;
41.38 (39.30) &\;\;\;\;
44.83 (41.28) \\ 
$A_{4,\text{reg}}(\rho)/\text{MeV}$ &\;\;\;\;
-0.0 (-0.0) &\;\;\;\;
-0.2 (-0.2) &\;\;\;\;
-0.3 (-0.6) &\;\;\;\;
-0.7 (-1.1) &\;\;\;\;
-1.0 (-1.4) &\;\;\;\;
-1.4 (-1.8) \\  
$A_{4,\text{log}}(\rho)/\text{MeV}$ &\;\;\;\;
0.4 (0.4)  &\;\;\;\;
0.8 (0.8) &\;\;\;\;
1.3 (1.1) &\;\;\;\;
1.5 (1.3) &\;\;\;\;
2.0 (1.8) &\;\;\;\;
2.4 (1.9)\\ 
 \hline \hline 
\end{tabular}
\end{minipage}
\begin{minipage}{17.8cm}
\caption
{Extracted values of $A_{4,\text{log}}(\rho)$ and $A_{4,\text{reg}}(\rho)$. The numbers in front (brackets) correspond to n3lo414 (n3lo450). The statistical errors (with respect to stepsize and grid length variations) of the results are of order $\pm0.1\,\text{MeV}$. For comparison we also show the results for $A_{2}(T=0,\rho)$. Note that $A_{4,\text{reg}}(\rho)$ includes the Hartree-Fock level results for $A_4(T=0,\rho)$.}
\label{tablelog}
\end{minipage}
\end{center}
\end{table*}

\begin{figure}[t] 
\centering 
\vspace*{-0.85cm}
\hspace*{-0.0cm}
\includegraphics[width=0.48\textwidth]{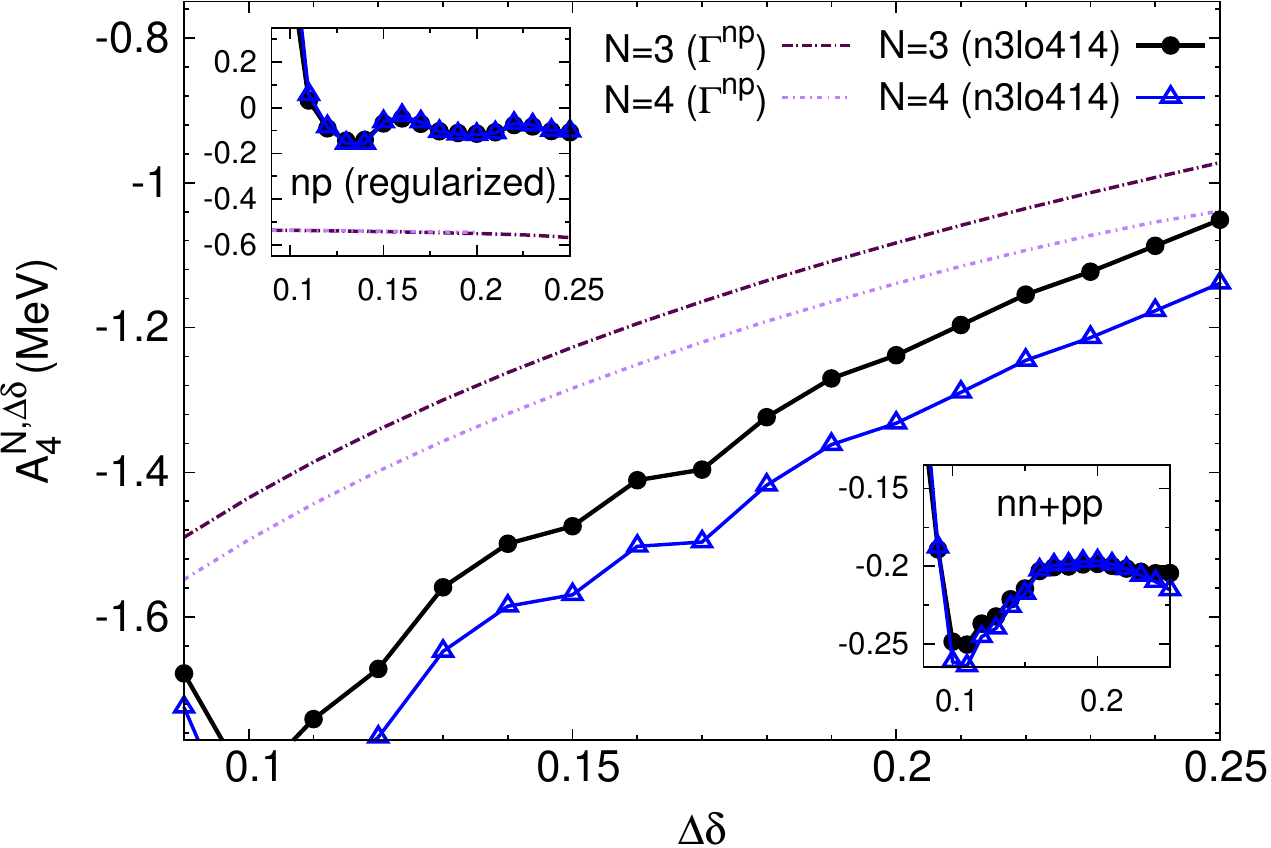} 
\vspace*{-0.7cm}
\caption{(Color online) Same as Fig.\ \ref{plotlog2} but for the quartic finite differences.}
\vspace*{-0.325cm}
\label{plotlog1}
\end{figure}

\begin{figure}[t] 
\centering 
\vspace*{-0.85cm}
\hspace*{-0.0cm}
\includegraphics[width=0.48\textwidth]{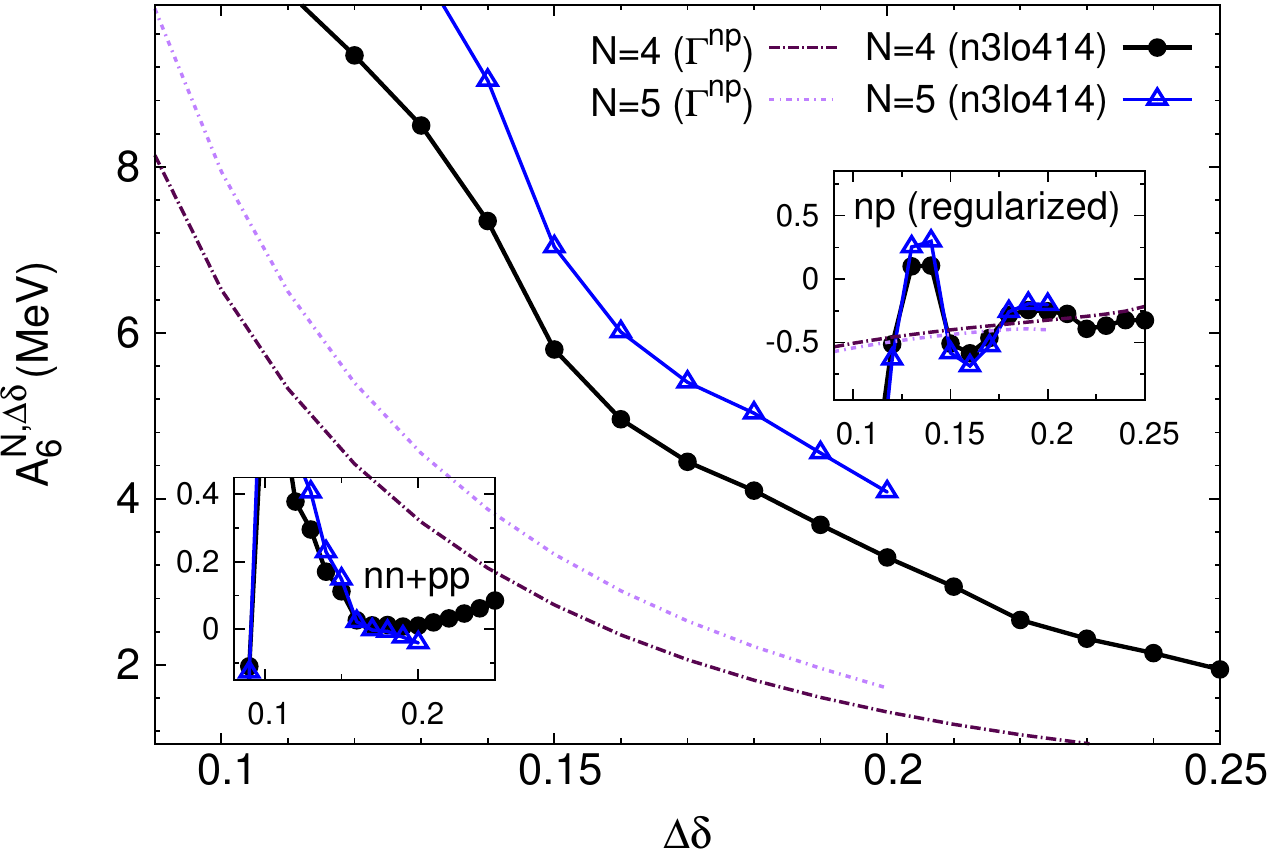} 
\vspace*{-0.7cm}
\caption{(Color online) Main plot: sextic finite-difference results for the np-channel second-order NN contribution at zero temperature (n3lo414, $\rho=0.20\,\text{fm}^{-3}$) and for the (suitably scaled) expression given by Eq.\ (\ref{Gammanp}). Top inset: analogous results for the np-channel contribution with the logarithmic term subtracted. Bottom inset: (combined) results for the nn- and pp-channel contributions.}
\vspace*{-0.525cm}
\label{plotlog2}
\end{figure}

\textit{Extraction of logarithmic term for chiral interactions.}
Eq.\ (\ref{T0expansion}) results in a stepsize dependence of the following form in the quartic and sextic finite-difference formulas:
\begin{align} \label{FDAT01}
A^{N,\Delta\delta}_{4}=&{A}_{4,\text{reg}}
+C^4_1(N) {A}_{4,\text{log}} 
+{A}_{4,\text{log}} \ln(\Delta\delta) 
\breq &
+C^4_2(N) {A}_{6,\text{log}} \Delta\delta^2 +\mathcal{O}(\Delta \delta^4), \\ \label{FDAT02}
{A}^{N,\Delta\delta}_{6}=&{A}_{6,\text{reg}}
+C^6_1(N) {A}_{4,\text{log}} \Delta\delta^{-2}
+{A}_{6,\text{log}} \ln(\Delta\delta)
\breq &
+C^6_2(N) {A}_{6,\text{log}}  +\mathcal{O}(\Delta \delta^2),
\end{align}
where the numbers $C^{2n}_i(N)$ are determined by the respective finite-difference coefficients $\omega_{2n}^{N,k}$.
From Eqs.\ (\ref{FDAT01}) and (\ref{FDAT02}) the leading logarithmic term is given by
\begin{align} \label{FDAT0diff1}
\Xi_4:= \frac{{A}^{N_1,\Delta\delta}_{4}-{A}^{N_2,\Delta\delta}_{4} }{C^4_1(N_1)-C^4_1(N_2)}
\simeq \,{A}_{4,\text{log}} , \\    \label{FDAT0diff2}
\Xi_6:=
\frac{{A}^{N_1,\Delta\delta}_{6}-{A}^{N_2,\Delta\delta}_{6} }{C^6_1(N_1)-C^6_1(N_2)}
\Delta\delta^2 
\simeq \,{A}_{4,\text{log}},
\end{align}
where the leading correction is proportional to ${A}_{6,\text{log}}\Delta\delta^2$.
For the $S$-wave contact interaction, where ${A}_{4,\text{log}}/{A}_{6,\text{log}}\simeq 2.60$, Eqs.\ (\ref{FDAT0diff1}) and (\ref{FDAT0diff2}) reproduce the exact value of ${A}_{4,\text{log}}$ to high accuracy. The n3lo414 results for ${A}^{N,\Delta\delta}_{4,6}$ are shown for grid lengths $N=3,4,5$ in Figs.\ \ref{plotlog1} and \ref{plotlog2}. For sufficiently large stepsizes the np-channel results for $A_4^{N,\Delta\delta}$ and $A_6^{N,\Delta\delta}$ exhibit the logarithmic and inverse quadratic stepsize dependence, respectively, expected from Eqs.\ (\ref{FDAT01}) and (\ref{FDAT02}). As expected from the analytic results for the $S$-wave contact interaction, this feature is absent for the nn- and pp-channel contributions. 

Averaging the values obtained from $\Xi_4$ and $\Xi_6$ as well as different stepsizes $\Delta \delta$ and grid lengths $N_1$ and $N_2$, the results for $A_{4,\text{log}}(\rho)$ are displayed in Table \ref{tablelog}.
Using the results obtained for $A_{4,\text{log}}(\rho)$ we then compute for both the chiral interactions and the spin-triplet ($\sim a_t$) $S$-wave contact interaction the quartic and sextic finite differences corresponding to
\begin{align}
F_{2,\text{NN}}^\text{regularized}(0,\rho,\delta):=&F_{2,\text{NN}}(0,\rho,\delta)
- A_{4,\text{log}}(\rho)\, \delta^4\, \ln|\delta|.
\end{align}
The results are plotted in the upper insets of Figs.\ \ref{plotlog1} and \ref{plotlog2}.
One sees that for the regularized second-order term the stepsize and grid length dependence of the finite differences is removed, and for sufficiently large stepsizes $A_4^{N,\Delta \delta}\simeq{A}_{4,\text{reg}}$ is approximately constant. Adding the Hartree-Fock results for $A_4(T=0,\rho)$, the extracted values for ${A}_{4,\text{reg}}(\rho)$ are given in Table \ref{tablelog}. 
In addition, we have extracted $A_{4,\text{reg}}$ and $A_{4,\text{log}}$ also by fitting the coefficients $a_\text{reg}$ and $a_\text{log}$ of a function $f_N(\Delta \delta)=a_\text{reg}+a_\text{log}[C^4_1(N)+\ln(\Delta \delta)]$ to the unregularized results for $A_4^{N,\Delta\delta}$, cf.\ Eq.\ (\ref{FDAT01}); the results for $A_{4,\text{reg}}(\rho)$ and $A_{4,\text{log}}(\rho)$ obtained in that way were found to match those given in Table \ref{tablelog}. 
Note that the relative size of ${A}_{4,\text{reg}}$ is smaller for the chiral interactions as compared to the spin-triplet $S$-wave contact interaction, where ${A}_{4,\text{reg}}/{A}_{4,\text{log}}=[3-60\ln(3)+4\ln(2)]/60\simeq -1.002$.

For clarification, we emphasize that the singularity of $A_{2n\geq4}$ at zero temperature is not a consequence of the presence of a contact interaction at second order, but a generic feature of many-body perturbation theory. 
The relevance of the $S$-wave contact interaction is that in that case the second-order term can be given in closed form. 
We have computed the second-order term also for model interactions of the one-boson exchange kind \cite{PhysRevC.87.014338}, and have found that the contributions from the nn- and pp-channels are regular, but the np-channel contribution to $A_{2n\geq4}$ is singular (with the finite differences exhibiting approximately logarithmic and inverse quadratic stepsize dependence, respectively). 
Incidentally, we note that in Ref.\ \cite{PhysRevC.91.065201} it was shown that the contribution to the quartic coefficient from iterated one-pion exchange is singular.

\begin{figure*} 
\centering 
\vspace*{-0.2cm}
\hspace*{-0.0cm}
\includegraphics[width=0.98\textwidth]{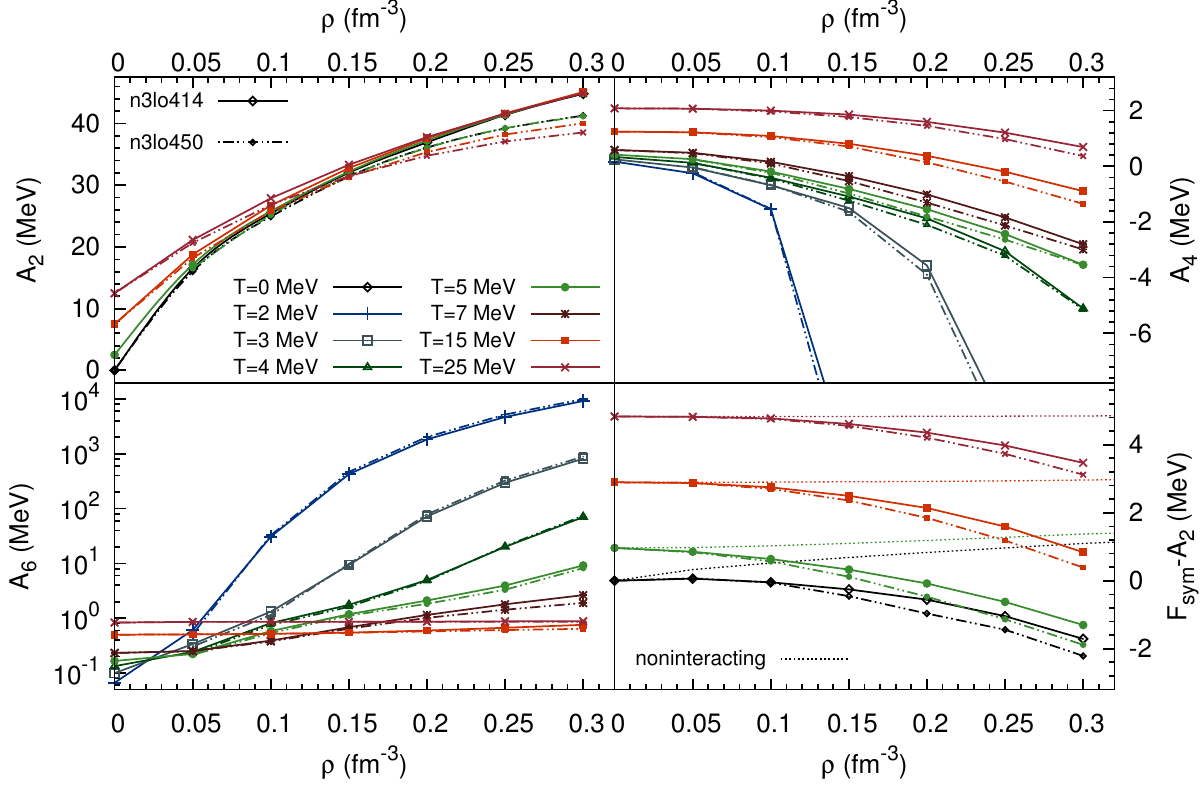} 
\vspace*{-0.2cm}
\caption{(Color online) Results for the quadratic, quartic and sextic Maclaurin coefficients for different densities and temperatures. Also shown is the difference between quadratic coefficient and the symmetry free energy (the dotted lines correspond to the results for a noninteracting nucleon gas).}
\label{plotmain1}
\end{figure*}

\section{Threshold for convergence of the Maclaurin expansion}\label{sec1}

The numerical results for the Maclaurin coefficients $A_{2,4,6}(T,\rho)$ of the EoS of infinite homogeneous nuclear matter obtained from the sets of chiral two- and three-body interactions n3lo414 and n3lo450 are displayed in Fig.\ \ref{plotmain1}. Only the dominant contributions at second order in perturbation theory are included.
Also shown is the difference between the quadratic coefficients $A_2(T,\rho)$ and the symmetry free energy ${F}_{\text{sym}}(T,\rho)={F}(T,\rho,1)-{F}(T,\rho,0)$. Overall, the results from n3lo414 and n3lo450 are very similar; the largest differences occur for the quadratic coefficient $A_2$ at high densities. The deviations between ${F}_{\text{sym}}$ and $A_2$ increase with temperature, but whereas for a noninteracting nucleon gas ${F}_{\text{sym}}- A_2$ (slightly) increases with density, for the interacting system ${F}_{\text{sym}}-A_2$ decreases with density. This difference in behavior is
predominantly from the first-order contribution from three-nucleon forces. Compared to the results for ${F}_{\text{sym}}(T,\rho)$ from Ref.\ \cite{paper2} the temperature dependence of  $A_2(T,\rho)$ is decreased in the case of n3lo414 and inverted for n3lo450 (at densities above saturation density).

\begin{figure}[t] 
\centering 
\vspace*{-0.0cm}
\hspace*{-0.0cm}
\includegraphics[width=0.48\textwidth]{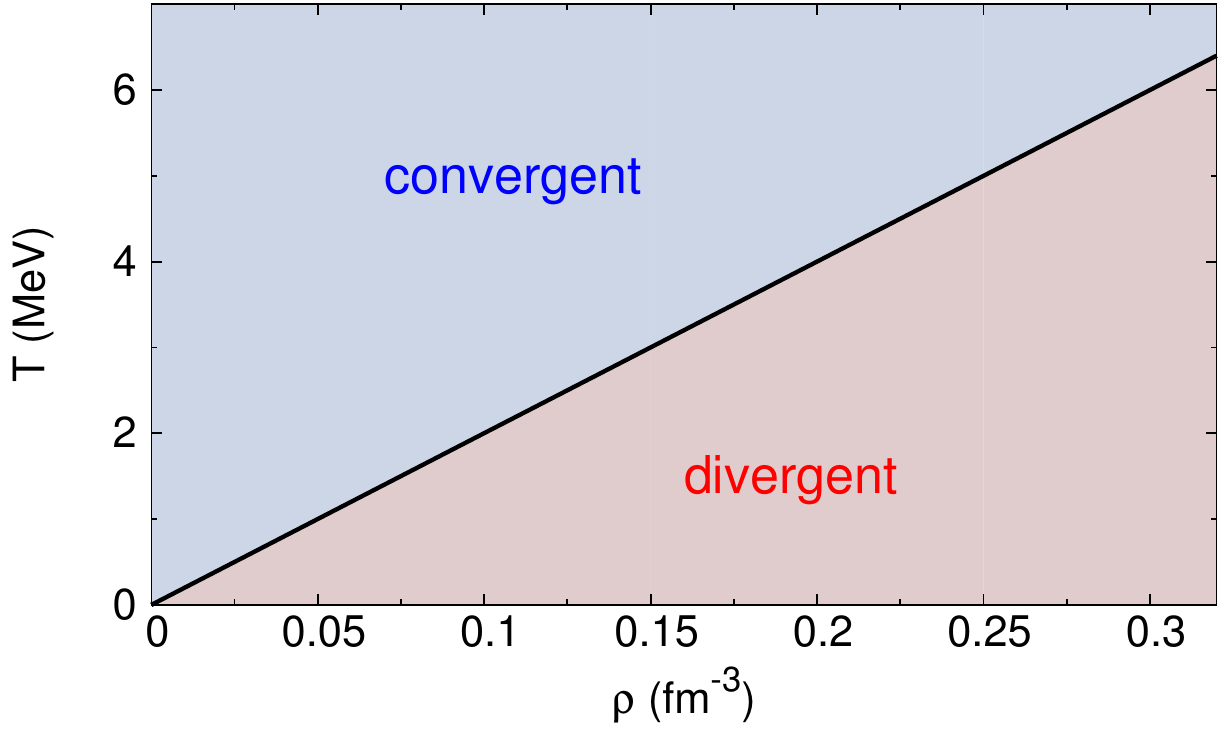} 
\vspace*{-0.5cm}
\caption{(Color online) Threshold line for the convergence of the isospin-asymmetry expansion of the free energy per particle, see text for details.}
\vspace*{-0.4cm}
\label{plotmain2}
\end{figure}

\begin{figure*} 
\centering
\vspace*{-0.45cm}
\hspace*{-0.0cm}
\includegraphics[width=0.98\textwidth]{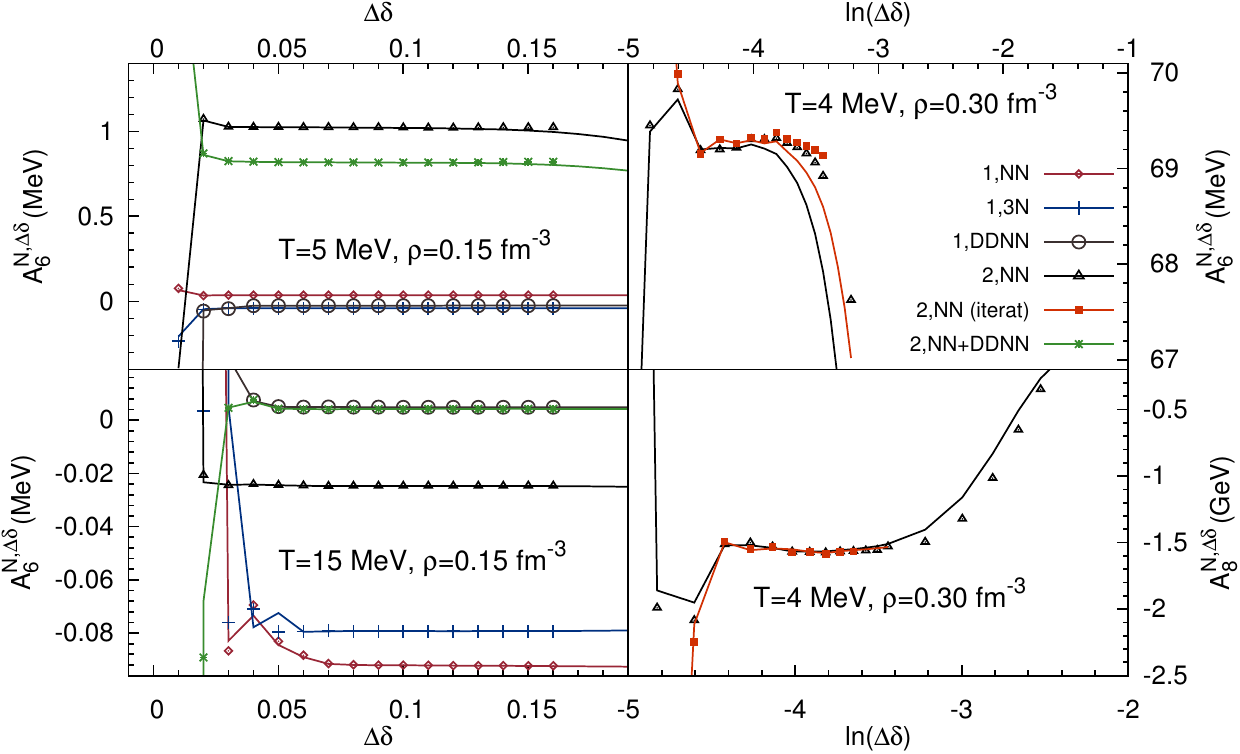} 
\vspace*{-0.2cm}
\caption{(Color online) Left column: finite-difference results for the first- and second-order (normal) contributions to ${A}_6$ at $T=5,15\,\text{MeV}$ and $\rho=0.15\,\text{fm}^{-3}$ (calculated using n3lo414). The lines correspond to $N=2+n$, the points to $N=3+n$; in each the stepsize variation extends from $\Delta \delta=0.01$ to $\Delta \delta=1/N$. At $T=15\,\text{MeV}$ the first-order NN and 3N contributions are given in units $10\,\text{keV}$. The approximate equality of the first-order DDNN and the second-order NN+DDNN results at $T=15\,\text{MeV}$ is coincidental.
Right column: ${A}^{N,\Delta \delta}_6$ and ${A}^{N,\Delta \delta}_8$ for the second-order NN contribution (np-channel only) at $T=4\,\text{MeV}$ and $\rho=0.30\,\text{fm}^{-3}$. Also shown are the results obtained from the iterative method, see Sec.\ \ref{sec23} for details.
Note that ${A}^{N,\Delta \delta}_8$ is given in units GeV.
}
\label{plotFDA1}
\end{figure*}

In the high-temperature and low-density region the Maclaurin coefficients obey $A_2> A_4> A_6(> A_8)$, with $|\xi-\zeta_6|<|\xi-\zeta_4|$.
This behavior breaks down when the temperature is decreased and the density is increased, leading to $A_2 \ll  A_4\ll A_6(\ll A_8)$ at high densities and low temperatures. 
In the sense that the expansion coefficients are hierarchically ordered at high temperatures and low densities the Maclaurin expansion with respect to the isospin-asymmetry can be rated as a convergent series in that regime, and as a divergent asymptotic series in the low-temperature and high-density region. To that effect, one can loosely identify a threshold line that separates the two regions. This line (roughly corresponding to $|\xi-\zeta_4|=|\xi-\zeta_6|$) is sketched in Fig.\ \ref{plotmain2}.  Note that since the divergent behavior is more pronounced for $A_{2(n+1)}$ than for $A_{2n}$ (cf.\ also Fig.\ \ref{plotFDA1}), one can expect that this threshold line rises when the isospin-asymmetry expansion is probed at increasing orders.

The strongly divergent behavior of the higher-order Maclaurin coefficients below the threshold line arises solely from the second-order many-body contribution. At the Hartree-Fock level the coefficients are hierarchically ordered $A_2> A_4> A_6(> A_8)$ also at high densities and low temperatures, and the Mauclaurin expansion is overall well-converged.
This explains why self-consistent mean-field theory calculations \cite{Carbone:2013cpa,PhysRevC.57.3488,Cai:2011zn,PhysRevC.89.028801,Chen:2009wv} have given small values of $A_4$ at zero temperature.

The question arises whether the nonanalyticity with respect to the isospin-asymmetry at low temperatures (i.e., the divergent behavior of the higher-order Maclaurin coefficients) is a genuine feature of the nuclear EoS or a feature that arises from probing higher-order derivatives of a perturbation series. It is however unlikely that this question can be resolved since nonperturbative approaches to the nuclear many-body problem presumably lack the precision for the numerical extraction of higher-order isospin-asymmetry derivatives. Quartic coefficients extracted from fits to nuclear masses have been relatively large, see Refs.\ \cite{PhysRevC.91.054302,PhysRevC.90.064303}, which could indicate that the isospin-asymmetry dependence of the (low-temperature) nuclear 
EoS is indeed nonanalytic.
More relevant for practical purposes is the question of the accuracy of the isospin-asymmetry parametrizations ${F}_{[2],[4],[6]}(T,\rho,\delta)$ at and beyond the leading quadratic order in the regions near and above the threshold line.

Representative results obtained for the different contributions to ${A}^{N,\Delta \delta}_{6}(T,\rho)$ and ${A}^{N,\Delta \delta}_8(T,\rho)$ are plotted in Fig.\ \ref{plotFDA1} for $T=4,5,15\,\text{MeV}$ and $\rho=0.15,0.30\,\text{fm}^{-3}$.
One sees that the numerical noise becomes visible only for very small values of $\Delta \delta$, and is more pronounced for larger temperatures. 
In the case of the first-order many-body contributions the results are well converged for a large region of $\Delta \delta$ values. At low temperatures the isospin-asymmetry dependence of the first-order DDNN contribution approximately matches that of the first-order contribution with genuine 3N forces, but this behavior deteriorates as the temperature is decreased.
For the Maclaurin coefficients corresponding to the second-order (normal) contributions at $T=15\,\text{MeV}$ and $\rho=0.15\,\text{fm}^{-3}$ the finite-difference method works also for large stepsizes, but for $T=5\,\text{MeV}$ a slight bending is observed as $\Delta \delta$ is increased. This bending gradually increases as the temperature is decreased and the density increased, cf.\ the results for ${A}^{N,\Delta \delta}_{6}$ and ${A}^{N,\Delta \delta}_8$ at $T=4\,\text{MeV}$ and $\rho=0.30\,\text{fm}^{-3}$. The cause of this bending is the large isospin-asymmetry slope (for low values of $\delta\neq 0$) and subsequent flattening of the higher-order isospin-asymmetry derivatives $\partial^n F/\partial \delta^n$, $n\geq 4$, in the low-temperature and high-density regime (this behavior can be inferred from evaluating the isospin-asymmetry derivatives of the function $\Gamma^{\text{np}}$ given by Eq.\ (\ref{Gammanp})).
If the stepsize is chosen too large, the behavior near $\delta=0$ is not resolved, which leads to the observed bending. 
However, as seen in Fig.\ \ref{plotFDA1}, even in that case a region where the stepsize dependence of the finite-difference results approximately vanishes can be found for small values of $\Delta\delta$. Note that in the region where the stepsize dependence nearly vanishes also the grid-length dependence is decreased.

\section{Isospin-asymmetry approximations}\label{sec3}

Here, we examine the accuracy of various isospin-asymmetry approximations with respect to the full isospin-asymmetry dependence of the free energy per particle $F(T,\rho,\delta)$. Overall, the accuracy of the leading-order quadratic approximation $F_{[2]}(T,\rho,\delta)=A_0(T,\rho)+A_2(T,\rho)\, \delta^2$ is determined by the magnitude of $F_\text{sym}(T,\rho)-A_2(T,\rho)$. In particular, it is $F(T,\rho,\delta)-F_{[2]}(T,\rho,\delta)\xrightarrow{\delta \rightarrow 1}F_\text{sym}(T,\rho)-A_2(T,\rho)$. From Fig.\ \ref{plotmain1} it can be inferred that the accuracy of $F_{[2]}(T,\rho,\delta)$ is decreased at the high temperatures probed in astrophysical simulations of core-collapse supernovae and binary neutron-star mergers.

\begin{figure}[t] 
\centering 
\vspace*{-0.0cm}
\hspace*{-0.0cm}
\includegraphics[width=0.48\textwidth]{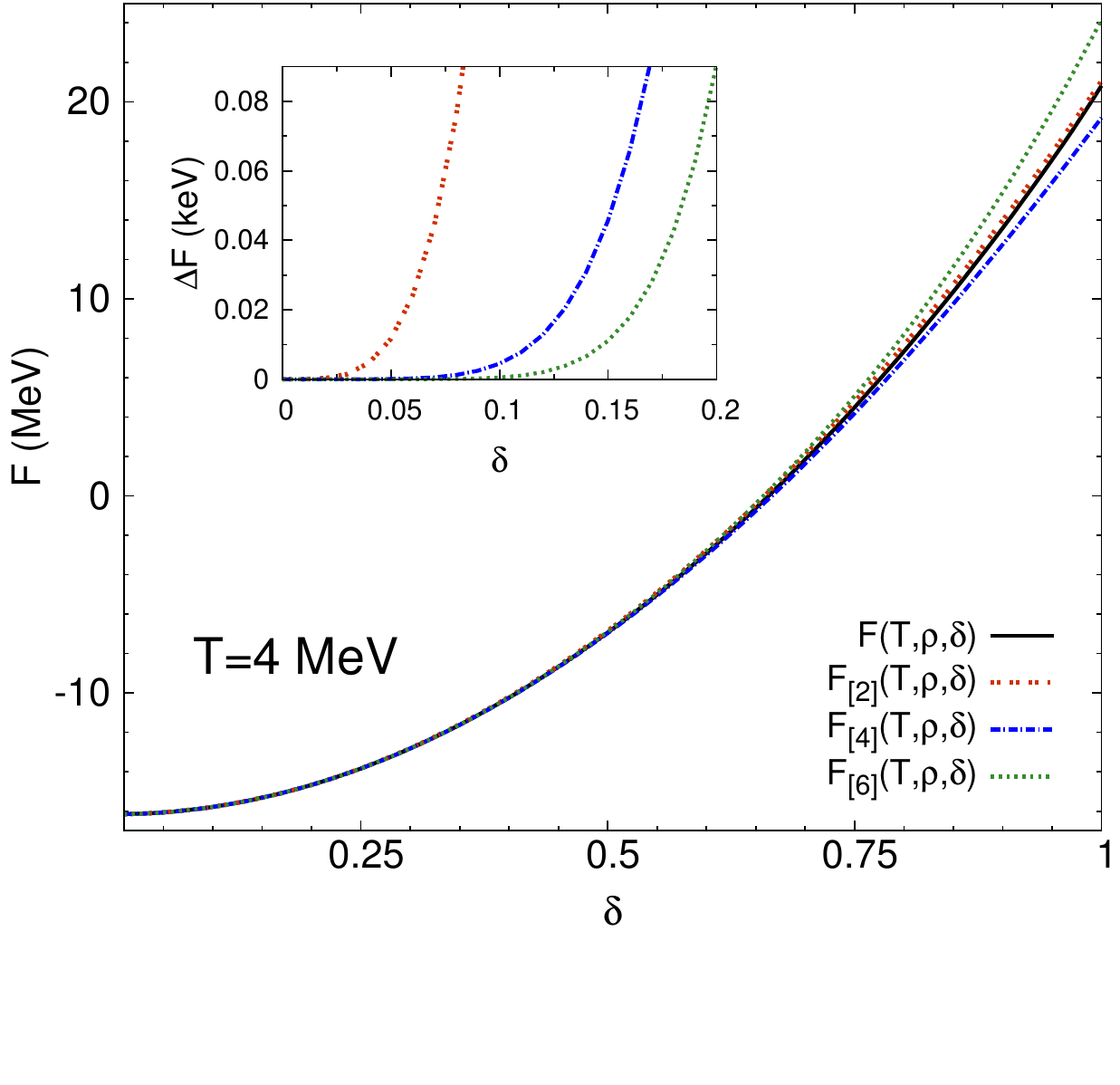} 
\vspace*{-1.8cm}
\caption{(Color online) Full isospin-asymmetry dependent free energy per particle $F(T,\rho,\delta)$ versus the quadratic, quartic and sextic polynomial isospin-asymmetry approximations at $T=4\,\text{MeV}$ and $\rho=0.20\,\text{fm}^{-3}$. The inset shows the deviation $\Delta F:=|F-F_{[2N]}|$ of the various approximations from the exact results for small neutron excesses $\delta\in[0,0.2]$ (results for n3lo414).}
\vspace*{-0.0cm}
\label{plotapprox1}
\end{figure}

The accuracy of the higher-order polynomial isospin-asymmetry parametrizations $F_{[4],[6]}(T,\rho,\delta)$ depends on the convergence behavior of the Maclaurin expansion roughly given by the threshold line in Fig.\ \ref{plotmain2}. At low densities and high temperatures including the quartic and sextic Maclaurin coefficients leads to an improved approximation also for large isospin asymmetries. For high densities or low temperatures the approximations $F_{[4],[6]}(T,\rho,\delta)$ improve upon $F_{[2]}(T,\rho,\delta)$ only for very small values of $\delta$. As a representative case we compare in Fig.\ \ref{plotapprox1} the isospin-asymmetry dependence of $F(T,\rho,\delta)$ with the one of the quadratic, quartic and sextic polynomial approximations ${F}_{[2],[4],[6]}(T,\rho,\delta)$ for $T=4\,\text{MeV}$ and $\rho=0.20\,\text{fm}^{-3}$ where $A_{2,4,6}/\text{MeV}\simeq(37.21, -1.87, 4.98)$. For small isospin asymmetries including the quartic and sextic coefficients leads to an improved approximation, but this behavior breaks down for intermediate values of $\delta$. In the neutron-rich regime the sextic and to a lesser extent also the quartic approximations deviate significantly from the exact result. The critical value of $\delta$ for which the accuracy of the higher-order approximations becomes inferior to the leading quadratic one decreases below the threshold line and vanishes (i.e., exceeds $\delta=1$) for temperatures and densities above it.

\vspace*{3mm}

\begin{figure}[t] 
\centering 
\vspace*{0.0cm}
\hspace*{-0.0cm}
\includegraphics[width=0.48\textwidth]{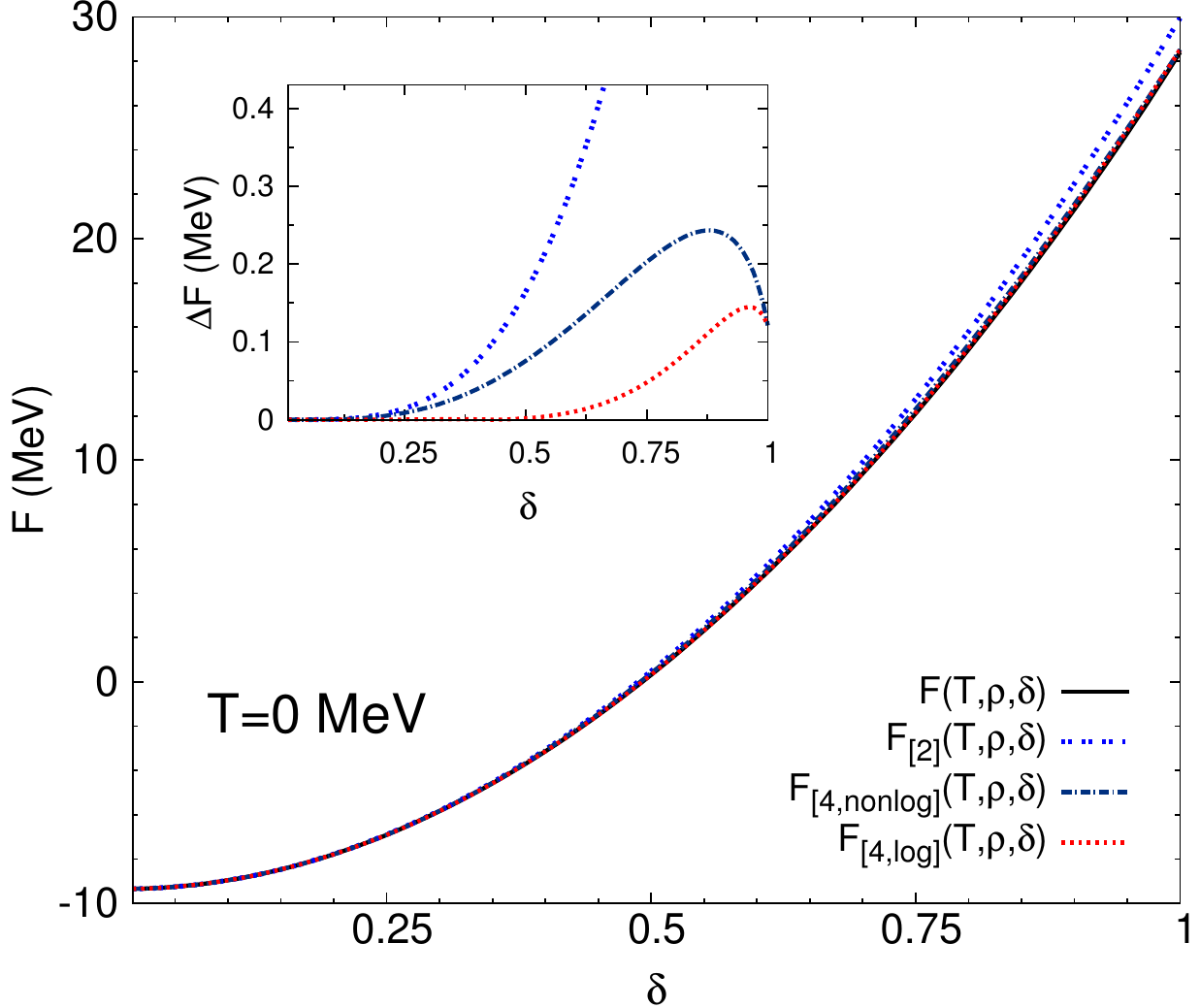} 
\vspace*{-0.7cm}
\caption{(Color online) Full isospin-asymmetry dependent free energy per particle $F(T,\rho,\delta)$ at $T=0$ and $\rho=0.25\,\text{fm}^{-3}$ versus the quadratic and quartic (with and without the logarithmic term) isospin-asymmetry approximations based on Eq.\ (\ref{T0expansion}). The inset shows the deviation of the various approximations from the exact results (results for n3lo450).}
\vspace*{-0.2cm}
\label{plotapprox2}
\end{figure}

In the zero-temperature case the following  approximation of the zero-temperature EoS can be constructed from the values for ${A}_\text{4,log}$ and ${A}_\text{4,reg}$ extracted in Sec.\ \ref{sec25}:
\begin{align} \label{Fapproxlog}
{F}_\text{[4,log]}(T=0,\rho,\delta)=&
{A}_0(0,\rho)+
{A}_2(0,\rho)\,\delta^2+
{A}_\text{4,reg}(\rho)\,\delta^4\breq &+
{A}_\text{4,log}(\rho)\,\delta^4 \ln|\delta|.
\end{align}
To identify the effect of the logarithmic term we consider also the quartic approximation of the zero-temperature EoS without the logarithmic term, i.e., 
\begin{align} \label{Fapproxnonlog}
{F}_\text{[4,nonlog]}(T=0,\rho,\delta)=&
{A}_0(0,\rho)+
{A}_2(0,\rho)\,\delta^2\breq &+
{A}_\text{4,reg}(\rho)\,\delta^4.
\end{align}
We emphasize that the higher-order approximations at zero temperature based on Eq.\ (\ref{T0expansion}) should be distinguished from those for the finite-temperature EoS based on the Maclaurin expansion Eq.\ (\ref{Fexpan1}).
The results obtained from Eqs.\ (\ref{Fapproxlog}) and (\ref{Fapproxnonlog}) are compared to the quadratic approximation and the exact results in Fig.\ \ref{plotapprox2}. One sees that including the logarithmic term considerably improves the description of the isospin-asymmetry dependence of the free energy per particle at zero temperature (this behavior deteriorates for densities $\rho \lesssim 0.5\rho_{\text{sat}}$ and $\rho \gtrsim 2\rho_{\text{sat}}$, where $\rho_\text{sat}\simeq 0.16\,\text{fm}^{-3}$). 

In principle, the determination of the Maclaurin coefficients via finite differences is equivalent to the (proper) fitting of the coefficients $B_{2n}$ of a polynomial function $F_\text{fit}^{M}(T,\rho,\delta)=A_0+\sum_{n=1}^{M} B_{2n} \delta^{2n}$ to the data $F(T,\rho,\delta)$. \footnote{The finite-difference weights $\omega_{2n}^{N,k}$ are in fact determined by the matching of Lagrange polynomials to the data, cf.\ Ref.\ \cite{coeff}.} However, as a consequence of the nonanalytic behavior of the isospin-asymmetry dependence in the low-temperature regime, the results for $B_{2n\geq 4}$ depend significantly on the fitting procedure (in that regime). In particular, in that regime a global unweighted (i.e., the data is distributed uniformly in the regime $\delta \in [0,1]$ and all data points carry equal weights) least-squares fit gives values for $B_{2n\geq 4}$ that deviate considerably from the Maclaurin coefficients $A_{2n\geq 4}$ (but the polynomial coefficients $B_{2n\geq 4}$ are in general still not hierarchical). By construction, the global fit gives a better description of the \textit{global} isospin-asymmetry dependence (as compared to the Maclaurin expansion), but the behavior in the region around $\delta=0$ is described less accurately. Crucially, the results for $B_{2n\geq 4}$ determined by a global fit are not unique, but depend strongly on the order of the fit polynomial, i.e., on the value of $M$ in $F_\text{fit}^{M}$. We have checked this behavior explicitly by fitting polynomials of order $M=2,\ldots,14$ to the data $F(T,\rho,k \Delta\delta)$, $k=0,\ldots,N$, with stepsize $\Delta \delta=0.01$ and grid length $N=100$. The global-fit results converge (with respect to increasing $M$) only in the high-temperature and low-density region, and in that case the values obtained for $B_{2n\geq 4}$ match the finite-difference results for the respective Maclaurin coefficients.

\section{Summary}\label{sec4}

In this work we have examined the isospin-asymmetry expansion of the free energy per particle of infinite homogeneous nuclear matter using second-order many-body perturbation theory with microscopic chiral low-momentum N3LO two-body and N2LO three-body interactions. The quadratic, quartic and sextic Maclaurin coefficients $A_{2,4,6}(T,\rho)$ corresponding to the expansion of the various many-body contributions have been extracted numerically using finite-difference methods. It has been found that the higher-order Maclaurin coefficients $A_{2n\geq 4}(T,\rho)$ are hierarchically ordered at high temperatures (and low densities), but diverge with alternating sign (for even and odd values of $n$) and increasing order of divergence for increasing values of $n$ in the zero-temperature limit.
As a consequence, in the low-temperature regime the free energy per particle is a nonanalytic (smooth for $T\neq 0$) function of the isospin asymmetry $\delta$, i.e., the Maclaurin expansion in terms of $\delta$ constitutes a divergent asymptotic expansion in that regime.
The divergent behavior is caused solely by the np-channel second-order contribution in the many-body expansion and is therefore not visible in (self-consistent) mean-field calculations of the nuclear equation of state. 
In the low-temperature and high-density region, the quartic and sextic expansion polynomials improve upon the leading-order quadratic approximation only for very small values of $\delta$, and fail in the neutron-rich regime.
The nonanalyticity of the isospin-asymmetry dependence in the low-temperature region entails that in that region there is no proper way of defining a global polynomial isospin-asymmetry parametrization beyond the quadratic one (cf.\ the last paragraph of Sec.\ \ref{sec3}).

The accuracy of the leading-order quadratic isospin-asymmetry approximation is inversely related to the magnitude of the difference between the quadratic Maclaurin coefficient $A_2(T,\rho)$ and the symmetry free energy $F_\text{sym}(T,\rho)=F(T,\rho,\delta=1)- F(T,\rho,\delta=0)$. The quantity $F_\text{sym}-A_2$ is significantly increased for the high temperatures relevant for astrophysical simulations of core-collapse supernovae and receives its main contributions from the noninteracting term and the first-order contribution from chiral three-nucleon forces, with the respective contributions carrying opposite signs (at high temperatures or low densities the noninteracting contribution dominates). Because the convergence rate of the expansion of the noninteracting contribution decreases significantly with temperature (in contrast to the behavior of the interaction contributions), a better approximation in the high-temperature regime is to calculate the noninteracting term exactly, and employ polynomial isospin-asymmetry parametrizations only for the interaction contributions, cf.\ Ref.\ \cite{paper2}.

Based on the analytic results for an $S$-wave contact interaction to second order, it was demonstrated that the leading singular term at zero temperature is logarithmic, $\sim \delta^4 \ln|\delta|$, also in the case of realistic chiral nuclear interactions.
The logarithmic term has been extracted from the analysis of linear combinations of finite differences, and it was found that the inclusion of this term leads to a significantly improved description of the isospin-asymmetry dependence of the zero-temperature equation of state. Future research will adress the influence of the logarithmic term on properties of neutron stars.

For a finite system the higher-order Maclaurin coefficients are finite at zero temperature, but diverge in the continuum limit. Hence, the higher-order perturbative contributions are still nonanalytic functions of the isospin asymmetry for sufficiently large particle numbers (analogous to the behavior of the infinite system at low temperatures).
Similar considerations apply for the case of infinite nuclear matter with  a pairing gap.
The question remains whether this nonanalyticity with respect to the isospin-asymmetry arises as a feature of perturbation theory only or is a genuine feature of the nuclear equation of state.

For neutron-rich matter it may be useful to consider an alternative parametrization of the isospin-asymmetry dependence in the form 
of an expansion in the proton fraction $Y_\text{p}=(1-\delta)/2$. 
Work along this line is in progress.

\vspace*{5mm}

\acknowledgements
We thank W. Weise, S. Petschauer, C. Drischler and A. Carbone for discussions, and the referee for useful comments.
This work is supported in part by the DFG and NSFC (CRC 110), and the US DOE Grant No.\ DE-FG02-97ER-41014.

\bibliographystyle{apsrev4-1}		
\bibliography{refs2}		

\end {document}